\documentclass{article}

\usepackage[utf8]{inputenc}

\usepackage{geometry}
\usepackage{graphicx}

\usepackage{animate}

\usepackage{amsmath}

\newcommand{\fsa}[1]{\left\langle #1 \right\rangle}

\usepackage[style=numeric, sorting=none]{biblatex}
\title{FUSE (Fusion Synthesis Engine): A Next Generation Framework for Integrated Design of Fusion Pilot Plants}

\author{
O. Meneghini$^{1,*}$, T. Slendebroek$^1$, B.C. Lyons$^1$, K. McLaughlin$^1$,\\
J. McClenaghan$^1$, L. Stagner$^1$, J. Harvey$^1$, T.F. Neiser$^1$,\\
A. Ghiozzi$^1$, G. Dose$^1$, J. Guterl$^1$, A. Zalzali$^1$, T. Cote$^1$,\\
N. Shi$^1$, D. Weisberg$^1$, S.P. Smith$^1$, B.A. Grierson$^1$, J. Candy$^1$\\
\\
$^1$General Atomics, San Diego, CA, USA\\
\\
$^*$Corresponding author: meneghini@fusion.gat.com
}

\date{\today}

\begin{document}

\maketitle

\begin{abstract}
The Fusion Synthesis Engine (FUSE) is a state-of-the-art software suite designed to revolutionize fusion power plant design. FUSE integrates first-principle models, machine learning, and reduced models into a unified framework, enabling comprehensive simulations that go beyond traditional 0D systems studies. FUSE's modular structure supports a hierarchy of model fidelities, from steady-state to time-dependent simulations, allowing for both pre-conceptual design and operational scenario development. This framework accelerates the design process by enabling self-consistent solutions across physics, engineering, and control systems, minimizing the need for iterative expert evaluations. Leveraging modern software practices and parallel computing, FUSE also provides multi-objective optimization, balancing cost, efficiency, and operational constraints. Developed in Julia, FUSE is fully open-source under the Apache 2.0 license, promoting transparency and collaboration within the fusion research community.
\end{abstract}

%=======
%=======
\section{Introduction}
%=======
%=======
Future fusion power plants will need to function in conditions vastly different from our current technological experiences. The 2022 Bold Decadal Vision underscored the necessity of economical fusion power, emphasizing the importance of effective modeling tools that provide actionable insights \cite{Hsu2023}. The ITER project underscored the necessity of early and iterative integration of nuclear analysis into the design phase, with late-stage design changes resulting in costly schedule adjustments \cite{titus2013}. The National Academies of Science, Engineering and Medicine highlighted \cite{Hawryluk2021}, fusion power plants need to progressively show robust heat removal, controlled material erosion, and minimal tritium loss over environmental cycles. This makes the evaluation of aspects like neutron shielding, tritium management, and materials lifetime and activation crucial for both safety and economic considerations. In this context, a whole facility modeling capability that can predict and assess the performance of fusion power plant concepts becomes indispensable.

Until recently the prevailing fusion power plant design processes rely excessively on system codes (like the GA Systems Code GASC \cite{Stambaugh2011}, ARIES systems code \cite{dragojlovic2010}, KAERI system code \cite{hong2008}, JAERI system code \cite{nakamura2012}, PROCESS \cite{franza2015}), which use simplified analytical models for the various parts of the plasma and plants. These codes are used to optimize key parameters such as cost and net power generation. However, the broad overview they offer belies the complexities of the individual systems, which are then probed further using high-fidelity simulations \cite{mao2019, coleman2019, Buttery2022, weisberg2023b, muldrew2024}. That's when discrepancies emerge. Despite the detailed insights of the high-fidelity models, their outcomes often clash due to the inherent simplifications of the initial models. The task then becomes reconciling these inconsistencies, requiring manual iteration to achieve a cohesive solution across all systems. This iterative method, involving multiple specialized teams, is not only resource-intensive but leaves lingering uncertainties, since even after arriving at this integrated solution, we're faced with a lingering question — how optimal is our result? The current approach tends to be more reactionary than strategic. Designers typically seek incremental enhancements within individual systems, only resorting to broader re-integration when one system undergoes significant alterations. This piecemeal approach lacks the systematic rigor needed to ensure a truly optimal integrated design.

To this end, the connections between the system and design codes can be consolidated by system analysis tool of intermediate (or even variable) fidelity, which can impose more consistency. Examples of such codes are SYCOMORE \cite{reux2015}, BLUEPRINT \cite{coleman2019}, MIRA \cite{franza2022}. This approach allows scoping and parametrizing multiple power plant configurations with a more consolidated physics and engineering modeling representation at a component level, but still keeping a holistic view of the entire fusion plant. Thus, providing margins for design improvements to a higher degree. Worth mentioning are also other efforts, such as the FREDA SciDAC project \cite{badalassi2023} which aims at coupling cutting edge high-fidelity solvers into a single multiphysics simulation environment of unprecedented accuracy, of course at the cost of increased computational complexity.

%=======
%=======
\section{FUSE framework}
%=======
%=======
General Atomics' (GA) commitment to the swift and assured deployment of fusion power is anchored in their utilization of cutting-edge integrated modelling tools \cite{GAFPP}. A testament to this is GA’s Fusion Synthesis Engine (FUSE) - a state-of-the-art software suite that seamlessly combines first-principle models, machine-learning techniques, and reduced models to facilitate comprehensive facility simulations that transcend the fidelity of traditional 0D systems studies. A product of GA's expertise in fusion theory and integrated simulations \cite{Meneghini2015, Meneghini2020, Slendebroek2023, Lyons2023}, FUSE was built from the ground up to achieve a series of objectives.
\begin{itemize}
\item To be modular, serving as a platform where various models to be integrated to yield power plant designs with consistent physics, engineering, control mechanisms, cost analyses, and balance of plant systems. 
\item To support a hierarchy of models fidelity (including machine-learning accelerated surrogates) for each of these topical areas.
\item To handle both stationary and time-dependent simulations, thus delivering power plant pre-conceptual designs, but also to develop operational scenarios, as well as a platform to eventually develop control systems and perform data analyses.
\item To be fast, and optimized to harness the power of parallelism and HPC systems to enable optimization studies, ensemble, and uncertainty quantification analyses.
\end{itemize}

\begin{figure}[!ht]
    \centering
    \includegraphics[width=\textwidth]{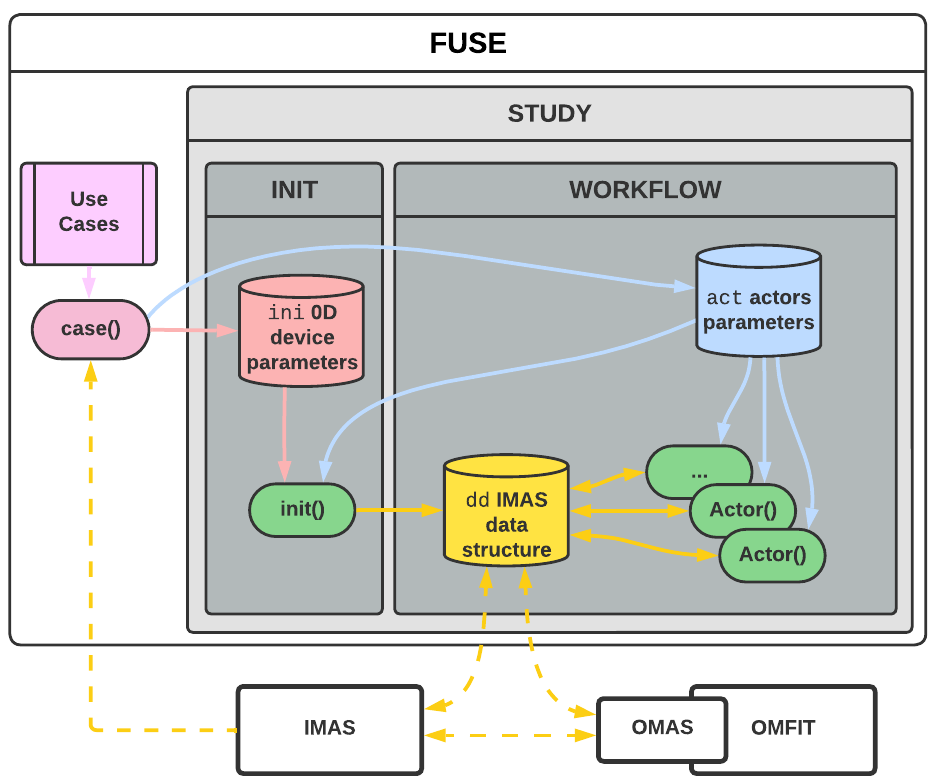}
    \caption{Overview of the software organization of the FUSE framework. Data is stored in the {\tt dd} data structure, which is based on the ITER IMAS ontology. Physics and engineering actors are the fundamental building blocks of FUSE simulations. Actors functionality is controlled via {\tt act} parameters. The data structure can be initialized starting from 0D {\tt ini} parameters. FUSE comes with a series of template use cases for different machines. Workflows perform self-contained studies/optimizations. FUSE can interface with the existing GA ecosystem of modeling codes (OMFIT/OMAS) as well as IMAS.}
    \label{fig:FUSE_overview}
\end{figure}

FUSE is a framework which allows its users to develop custom modeling \emph{workflows} tailored on their specific research and design objectives. These can range from the intricate design of singular components to the holistic optimization of an entire power-plant concept, and even the comparative analysis of varying power-plant concepts. FUSE accomplishes this by adopting the modular design shown in Fig.~\ref{fig:FUSE_overview}. At the core of FUSE simulations are the physics and engineering \emph{actors}, the functionality of which is controlled by specific {\tt act} parameters. All actors operate on a centralized data structure {\tt dd} that is rooted in the ITER Integrated Modeling and Analysis Suite (IMAS) \cite{Imbeaux2015} ontology. FUSE simulations can be initialized by either placing data directly in {\tt dd} (for example when data is already available in IMAS format) or by triggering an \emph{init} procedure that populates {\tt dd} starting from elementary (0D) parameters defined in the {\tt ini} structure. FUSE provides an broad range of \emph{use-cases} (ITER, ARC, DIII-D, and various other tokamak designs) that users can use as guide to setup new machine studies, and are also used for regression testing. FUSE's ability to write data to IMAS file formats enables its seamless integration with the broader fusion modeling ecosystem. These include the OMFIT \cite{Meneghini2015} integrated modeling framework, the OpenMC \cite{romano2015} neutronics software, the control systems environment TokSys \cite{walker2015}, and the finite element analysis model GATM \cite{leuer2023}. FUSE can also directly interface with the IMAS environment (and the plethora of models compatible with it) by exchanging data via IMAS's native HDF5 file format.

%==========
\subsection{An open-source ecosystem for fusion written in Julia}
%==========
To achieve its speed and scalability objectives FUSE was entirely written in the Julia programming language \cite{julia}. Julia is a Just In Time (JIT) programming language that combines the ease of use and familiar syntax of languages like Python, R, and Matlab with the performance of languages such as C, C++, or FORTRAN \cite{julia_benchmarks}. This means researchers and developers can prototype and deploy in the same environment, reducing the need to create models in one language and then convert them to a faster production language, thus saving time, reducing costs, and minimizing potential errors.

The high performance of Julia makes it possible to develop the framework infrastructure, as well as the physics and engineering models natively in the same language. There are many benefits to having everything being written in Julia, the most important being:
\begin{enumerate}
\item Natural tight coupling between models (ie. direct in-memory data exchanges) avoiding performance loss associated with writing and reading files
\item Support for automated differentiation (AD) for efficient computation of derivatives, which are essential for optimization, sensitivity analysis, and machine learning applications
\item The system simplifies the management and integration of various dependencies, significantly reducing the overhead typically associated with maintaining a consistent and operational environment
\end{enumerate}

Initiated as an internal R\&D project at General Atomics \cite{GAMFE} in 2021, the FUSE project is now licensed under Apache 2.0 \cite{apache2} and hosted on GitHub in the ProjectTorreyPines organization \cite{ptp}. It is installed via the top-level {\tt FUSE.jl} package, part of an ecosystem comprising nearly 30 Julia packages. All physics and engineering models are coupled to FUSE through the {\tt IMAS.jl} package, which defines the common ontology for model integration. Documentation is available online \cite{fuse_help}.

%==========
\subsection{An extended IMAS ontology}
%==========
The ITER Integrated Modeling and Analysis Suite (IMAS), is underpinned by the ontology known as the ITER Physics Data Model (PDM), which structures data around nearly 80 hierarchically ordered Interface Data Structures (IDSs). These IDSs span various modeling topics like equilibrium, kinetic profiles, and sources, as well as experimental areas including magnets, diagnostics, and heating systems. Each IDS encapsulates the necessary data pertaining to its associated plasma or tokamak subsystem. Within an IDS, every quantity is meticulously defined, specifying units, coordinates, numerical type, and a comprehensive description. For facilitating code coupling and driving integrated modeling workflows, ITER relies on these standardized IDSs to ensure consistent data exchange across code components.

Similarly, in FUSE the modularity of the framework is enabled by leveraging a modified version of the ITER-PDM, which has been extended to include the description of sub-systems, such as the blanket or the balance of plant, that are not implemented in the original ontology (as of version 3.41.0). We refer to the extended ontology as the FUSE-PDM, and all FUSE models rigorously exchange data according to such specification. Data management per FUSE-PDM specifications is accomplished via the {\tt IMAS.jl} package, which draws from the GA team's experience with the OMAS library \cite{Meneghini2020}.

A distinguishing feature of {\tt IMAS.jl} is its ability to lazily evaluate derived quantities within the data structure. Such a system of dynamic expressions ensures consistency and provides an elegant solution to the \emph{mismatching-interfaces problem}, where preceding models might not furnish all the derived data needed by subsequent models. To ensure consistency throughout FUSE, {\tt IMAS.jl} directly incorporates a comprehensive set of mathematical, physics, and engineering routines. This eliminates the need for individual packages to re-implement these routines.

A standout feature of {\tt IMAS.jl} is its ability to easily manage IDSs that are non-homogeneous in time. This capability is crucial for facilitating the creation of comprehensive time-dependent simulations. The process is streamlined by introducing a concept of a \emph{global time}. When this global time of interest is defined, the data structure's API takes charge. Instead of the user having to manually decipher the time coordinate for each accessed element, the API determines it and accordingly interpolates (or updates, if writing) the data for that specific time. Moreover, for efficient management of extensive time series, {\tt IMAS.jl} employs a memory-saving strategy: it stores only the differences between consecutive time slices within an array of structures, rather than the entirety of each slice.

The {\tt IMAS.jl} package also retains the ability to interoperate with the original IMAS infrastructure by directly reading and writing HDF5 binary files using the native ``tensorized'' IMAS data format \cite{Meneghini2020}. In addition {\tt IMAS.jl} supports reading and writing data in the JSON ASCII format, which has proven to find broad adoption among different projects that use the OMAS Python library. The ability to I/O data in these data formats does not depend on either the original IMAS infrastructure nor OMAS being installed.

Last but not the least {\tt IMAS.jl} makes use of Julia's Plots.jl and uses multiple dispatching mechanism to provide contextual (and composable) plotting capabilities throughout the data structure. By defining plots at the level of the data structure, plotting is agnostic to the actor that populated the data structure.

%==========
\subsection{Balancing models fidelity and speed}
%==========
The framework modularity allows users to select from a range of models' fidelity for each of the sub-systems that compose a fusion power plant. At the time of writing, models' development part of the FUSE project has been aimed to capture the complex dynamics of the coupled systems, while enabling rapid design iterations. This has been achieved by striking a balance between the computational efficiency of simpler analytical models and the intricate detail and cost of execution of first-principles simulations.

In integrated simulations it is critical to maintain a consistent model fidelity across subsystems. For the majority of applications, having every model in the simulation harmoniously balance fidelity with computational speed is not just desirable, but often necessary. This uniformity assures consistent results, preventing the discrepancies that can arise when detailed outputs from high-fidelity models aren't effectively leveraged by their low-fidelity counterparts. It also promotes computational efficiency, preventing any singular, overly detailed component from becoming a bottleneck in the simulation process. The guiding principle behind model selection has been to have sufficient fidelity to get the interfaces between subsystems about right, so in-depth sub-system studies would not upend the whole design. While at this time the judgement about what models to develop and include in the simulations is based on experience of the field-experts and developers, in the future this could be more systematically addressed by comparing the statistical validation error of different models against the uncertainty resulting from the propagation of uncertainties through the different models.

\begin{figure}[ht]
    \centering
    \includegraphics[width=0.5\textwidth]{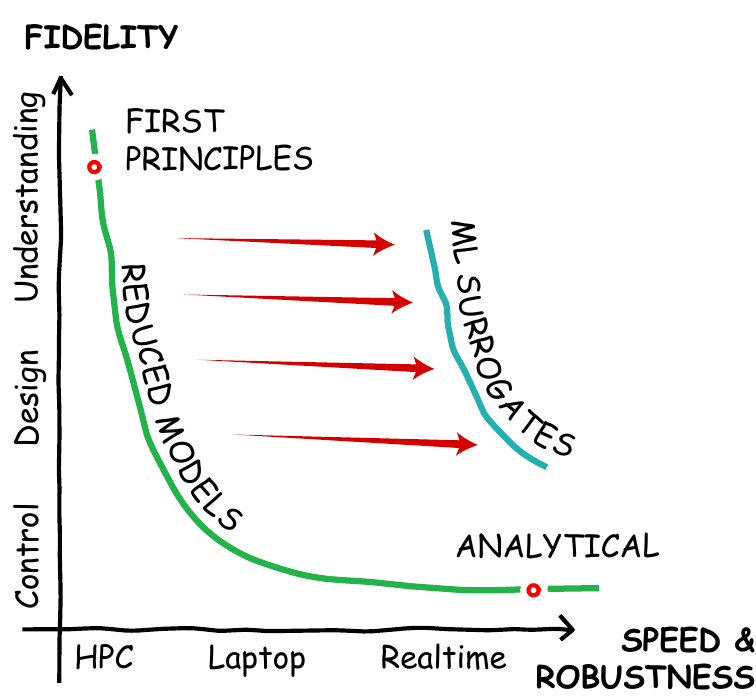}
    \caption{Schematic indicating the generally observed trade-off between model fidelity versus speed. In addition to being faster, reduced models also tend to be more robust. Machine learning models can be used to break such tradeoff.}
    \label{fig:speed_vs_accuracy_3}
\end{figure}

Whenever the fidelity requirements demand for models that would become a significant bottleneck in the FUSE workflows, the team has resorted to acceleration via machine learning (ML) (see Fig.~\ref{fig:speed_vs_accuracy_3}). ML surrogate models, trained on data from detailed simulations, predict system outputs at a remarkable speed. After training, they can rapidly produce results, several magnitudes faster than comprehensive simulations. This means the surrogate model, while maintaining the accuracy of its high-fidelity counterpart, can offer results with the agility of an analytical model. These ML surrogate models (also implemented in Julia) are used for rapid design iterations and optimization tasks.

By adopting Julia as a performant programming language, using tight-coupling between models, leaning into reduced and ML models, and performing aggressive profiling throughout its ecosystem, FUSE can perform a start-to-finish machine design within one or two minutes.

%==========
\subsection{Iterative workflow for models' refinement}
%==========
Following the execution of integrated simulations, the next phase involves individual teams diving deep into the detailed design of each sub-system. Starting from the outcomes of these simulations provides a consistent baseline, ensuring that all teams commence their design processes from a unified reference point. In practical terms, a fiducial design point is disseminated among various teams through the {\tt dd} data structure, saved in the JSON format \cite{Meneghini2020}. This format is easily accessible by a range of tools (see Fig.~\ref{fig:FUSE_overview}), including the integrated modeling framework OMFIT \cite{Meneghini2015}, the finite element analysis tool GA Tokamak Model (GATM) \cite{leuer2023}, the plasma control system platform TokSys \cite{humphreys2008}, and a specific 3D neutronics package \cite{weisberg2023b, grierson2023}. Furthermore, the capability to store {\tt dd} in the native IMAS HDF5 format offers seamless integration with any modeling tool that is IMAS-compatible.

As these teams of experts intensify their focus on specific sub-systems, discrepancies — albeit fewer than starting with just a systems code — inevitably surface. To effectively manage and address these discrepancies, a robust feedback mechanism is essential. Periodic checkpoints allow topical teams to share their findings. Deviations from the original baseline can be integrated back into FUSE either via explicit constraints to the simulations, or by improving the models used in the integrated simulations, be it adjustments, new integration, or incorporating data-driven enhancements like machine learning. As the design progresses, it's thus crucial to re-run integrated FUSE optimizations at intervals or after substantial design alterations. This not only provides an updated reference but also allows teams to assess the impact of their changes on the entire system.

The aim of this process is not only to obtain an optimal and feasible fusion power plant design, with all components working harmoniously, but also to continuously improve the models and their integration, which ever-evolving, form the backbone of the adaptive and robust integrated design framework that is FUSE.

%=======
%=======
\section{FUSE physics and engineering actors}
\label{sec:actors}
%=======
%=======
In the FUSE modeling framework, \emph{actors} serve as the fundamental building blocks for simulations in both physics and engineering domains. These actors exclusively manipulate data from the {\tt dd} structure and their functionality can be tuned using the associated {\tt act} parameters. In FUSE it is possible to combine multiple actors into other actors, enabling intricate and multi-faceted simulations.

FUSE's fidelity hierarchy is enabled by the concepts of \emph{generic} and \emph{specific} actors. While generic actors lay out the fundamental definitions related to a certain physics or engineering component, the specific actors offer precise model implementations for those defined components. For example, ActorEquilibrium is a generic actor, while ActorSolovev, ActorCHEASE, ActorTEQUILA are specific actors. By convention, {\tt act.Actor[Generic].model} is used to select the specific actor being used to carry out a calculation. All specific actors expect and fill the same entries in {\tt dd}, which is something that is facilitated by the use of {\tt IMAS.jl} expressions. To ensure modularity and future adaptability, actors calling other actors avoid hardcoding of specific actors whenever feasible in favor of generic actors.

FUSE provides broad range of physics and engineering actors, which we present grouped into four broad categories: plasma in Sec.~\ref{models:facility}, H\&CD in Sec.~\ref{models:hcd}, power plant in Sec.~\ref{models:bop}, and facility in Sec.~\ref{models:facility}. But first we should discuss how FUSE simulations are initialized from scalar quantities.

%==========
\subsection{Initialization of FUSE simulations}
%==========
The initialization of a multi-physics simulation can be a non-trivial task. For example, performing a 1.5D simulation of a tokamak plasmas generally requires a starting from an initial guess for the equilibrium solution, along with kinetic profiles that detail species densities, temperatures, and rotation, as well as source profiles that encompass heating, current drive, and torque.

Such starting data must be made available in the {\tt dd} data structure for the actors to be able to operate on it. While this could be done manually, the comprehensive and deeply nested hierarchical nature of {\tt dd} makes this approach tedious and error prone. To complicate things further, the initialization procedure is highly dependent on the particular study that one wants to carry out. For example, a particular study may call for starting from an experimental Grad-Shafranov equilibrium, using heating and current drive sources from existing simulations, and parametrize the kinetic profiles in terms of 0D parameters. Another study may already have all the starting data in IMAS format. Other studies will call for yet another set of unique initial conditions. While such flexibility is necessary, it inevitably adds complexity.

To address these challenges, FUSE has an initialization routine that is designed to provide a physically plausible (albeit not necessarily self-consistent) starting point. This approach is similar to the PRO-CREATE method outlined in Ref.~\cite{Slendebroek2023}, but it has been further expanded to encompass the initialization of the entire engineering spectrum of the plant, including construction, materials, and operational requirements. Importantly, all of the complexity of handling different user's entry-points into the FUSE simulation framework is relegated to the {\tt init()} procedure in FUSE, which allows actors to be truly agnostic of the workflow that is being run.

The behavior of the {\tt init()} procedure is controlled via a {\tt ini} input, which allows to specify if data should be taken from scalar quantities and/or data available in some file in IMAS format. Users that have data in other legacy formats, such as gEQDSK equilibrium files, U-files, and such, are encouraged to convert their data in IMAS format outside of FUSE, which can be done in a variety of ways, for example via OMFIT's fusion data files classes.

Parameters defined {\tt ini} and {\tt act} can be uncertain as well as time dependent functions. These features are used to define in an intuitive way uncertainty ranges in uncertainty propagation workflows, optimization ranges of design parameters in machine optimizations (see Sec.~\ref{sec:multi_objective_optimization}), actuators trajectories in time-dependent simulations (see Sec.~\ref{sec:timedep}).  The {\tt ini} and {\tt act} functionalites are implemented as part of the {\tt SimulationParameters.jl} package in the FUSE Julia ecosystem.

%==========
\subsection{Plasma}
\label{models:plasma}
%==========
Modeling the plasma behavior is at the heart of any Fusion Power Plant (FPP) simulation. In this respect FUSE can handle both the steady-state and the time-dependent characteristics of the plasma. Emphasis is placed on the coupled equilibrium and transport modeling of particles, energy, and momentum in the plasma core. Core-pedestal (or edge) coupling is also included in the simulation. For the time being SOL modeling is limited to onion-skin models. Separate analyses examine the operational limits, mainly implemented my analytical models.

%=============
\subsubsection{Equilibrium}
%=============
FUSE supports three different equilibrium models. The simplest implements an analytic extended-Solov'ev solution \cite{cerfon2010} to the Grad-Shafranov problem in which $p'$ and $FF'$ are taken to be constants (where $p$ is the pressure, $F$ is the toroidal field function, and prime represents derivation with respect to the poloidal magnetic flux $\Psi$). While fast, this model is not valid for realistic plasma configurations. To incorporate more complex pressure and current profiles, FUSE was originally coupled to the inverse Grad-Shafranov equilibrium solver CHEASE \cite{lutjens1996}. While CHEASE can still be sued, we have developed a Julia-based, fixed-boundary, inverse equilibrium solver named TEQUILA that is now the default in FUSE. Unique among equilibrium solvers, TEQUILA uses the Miller extended-harmonic (MXH) parametrization \cite{arbon2020} which provides an efficient representation of flux-surfaces with very few Fourier modes. TEQUILA is provided the MXH parametrization of the fixed plasma boundary, and then fixed profiles for either $p$ or $p'$ and either $FF'$, $\fsa{J_t/R}$, or $\fsa{J_t/R}/\fsa{R^{-1}}$, where $\fsa{\cdots}$ is a flux-surface average, $J_t$ is the toroidal current density, and $R$ is the major radius. These profiles can be held fixed with respect to the normalized poloidal flux or the normlized toroidal flux. TEQUILA uses the standard Picard iteration scheme with relaxation to solve the nonlinear Grad-Shafranov equation. $\Psi$ is expanded in cubic-Hermite finite elements in the radial flux coordinate ($\rho$) and Fourier modes in the poloidal angle ($\theta$). In each iteration, the toroidal elliptic operator is inverted. The MXH parameterization of each flux surface is then updated so as to eliminate the $\theta$-dependence of $\Psi$. The toroidal current density is updated based on the new $\Psi(\rho)$ values and fixed profiles provided. As TEQUILA converges, it solves for both $\Psi(\rho)$ and the shape of the flux surfaces consistent with the input profiles and boundary.

Fixed-boundary equilibrium solvers do not provide the poloidal field solution in the vacuum region, outside of the last closed flux surface. For this purpose, FUSE uses the Green’s function method to determine the poloidal-field coil currents needed to best produce the desired equilibrium shape. First we compute the magnetic flux generated from image currents on the plasma boundary in the fixed-boundary solution. A least-squares optimization is then performed to determine the poloidal-field coil currents that produce an equivalent flux to the image currents at a fixed number of points near the plasma boundary. The location of saddle points (X-points) and additional points outside the plasma on the separatrix (e.g., strike points) can also be prescribed. The free-bounday flux at any point (inside or outside the plasma can then be found as $\Psi_{free} = \Psi_{fixed} - \Psi_{image} + \Psi_{coils}$. This approach forms the basis for FUSE's optimization of the PF coils placement as described in Sec.~\ref{ActorPFcoilsOpt}.

Finally, we note that a pure-Julia implementation of a free-boundary equilibrium solver is under way, which will become openly available when ready for public release.

%=============
\subsubsection{Current evolution}
%=============
FUSE uses new current diffusion solver named QED (Q Evolution by Diffusion). Written in pure Julia, the code uses the Galerkin method with Hermite-cubic finite elements to advance a diffusion equation for the rotational transform. QED can be advanced time-dependently (using a theta-implicit algorithm) or solve for the steady-state current profile. Constant current and constant loop voltage boundary conditions have been implemented. QED has been benchmarked against the current-diffusion routines within the TRANSP \cite{transp} and ONETWO \cite{onetwo} codes and provides a lightweight alternative for evolving the current without the need for a full transport solver.

%=============
\subsubsection{Core transport and pedestal}
%=============
In order to study present fusion experiments, and plan future fusion reactors, it is important to accurately model the heat and particle losses due to turbulence in magnetically confined plasmas. For this, FUSE uses a flux-matching approach (as in the TGYRO code \cite{candy2009}) to find the stationary solution to the 1.5D transport problems for the electrons and ion temperatures, electrons and ion densities, and the rotation profiles. The fluxes are matched at a number of core radial locations, and the profiles are reconstructed with a boundary point at the pedestal appropriate for the conventional H-mode operational regime, assuming that the profiles inverse scale length varies linearly between those flux-matched points \cite{meneghini2017}.

Designed with modularity at its core, the FUSE transport solver (named FluxMatcher) allows for the incorporation of transport fluxes from models of varying fidelity, including TGLF~\cite{Staebler07, Staebler16, Staebler21} and the TGLF Neural-Net (TGLF-NN) \cite{meneghini2017} models for turbulent transport, and  NEO \cite{belli2008}, Chang-Hinton, and Hirshman-Sigmar for neoclassical transport. Of these TGLF and NEO are loosely coupled to FUSE. The pedestal model used the EPED-NN model. The TGLF-NN and EPED-NN models \cite{meneghini2017} have been both re-implemented in pure Julia.

In addition a pure-Julia implementation of TGLF, named {\tt TJLF.jl}\cite{neiser2024}, has also been developed. Benchmarks against the original FORTRAN implementation of TGLF shows the two codes matching within numerical precision. Julia's native multi-threading capabilities allow TJLF to take full advantage of modern CPUs with numerous cores. A recent implementation of a the Miller extended harmonic geometry was first prototyped in TJLF, and later ported to FORTRAN. Having access to the internals of the code allows re-using of some of the spectra information between iterations of the transport solver, allowing for faster calculations and accelerated convergence. Ongoing work focused on leveraging the Julia's auto-differentiation capabilities to efficiently evaluate the fluxes sensitivity to the input parameters, as well as on using native GPU compatibility to accelerate calculation of energetic particle transport (TGLF-EP).

The sources and sinks of energy include heating from the heating and current drive (H\&CD) actuators, the fusion source derived from the alpha slowing down distribution, as well as Brehmstralung, synchrotron, and line radiation. External actuators contribute to various channel sources (energy, particle, momentum) based on the specific type of actuators, whether it's neutral beam injectors (NBI), electron cyclotron (EC), lower hybrid (LH), ion cyclotron (IC), or pellets.

Time derivative sources for each of the transport channels are also included when running time-dependent simulations (see Sec.~\ref{sec:timedep}). These time derivative sources oppose rapid changes in the kinetic profiles, and allow for time evolution to be correctly modeled even when using a flux-matching approach. The resulting time-implicit transport modeling scheme allows FUSE to take large time steps without sacrificing accuracy.

%=============
\subsubsection{Stability and operational limits}
%=============
In FUSE, the evaluation of plasma stability is grounded in models that predominantly rely on simplified semi-empirical relationships. These models offer foundational guidance for stability targets, encompassing a spectrum of stability limits such as those related to $\beta$, current, density, and elongation. These limits are systematically organized into collections tailored for distinct plasma scenarios. Depending on the scenario, different stability limit models might be invoked. For instance, one might examine multiple forms of the same limit, such as a $\beta_{\rm n}$ limit influenced by shaping or a $\beta_{\rm n}$ limit steered by bootstrap current. While the selection of these scenarios currently rests with the user, the overarching ambition is to automate this process, allowing for model choices to be contingent on a given plasma state that should be identified based on different set of parameters.

Vertical stability is treated with a dedicated model. Any viable tokamak design must (passively) bring down the growth rate of the open-loop vertical instability to levels that are realistic for closed-loop control. Two important non-dimensional quantities characterizing vertical stability are the inductive stability margin \cite{portone2005, humphreys2009} and the normalized growth rate \cite{olofsson2022}. The inductive stability margin, $m_s$, is negative if the vertical growth rate is Alfvenic and positive if the growth rate is on the wall-penetration timescale. Alfvenic growth is uncontrollable, and thus one typically constrains $m_s > 0.15$ as a margin to avoid such ideal instabilities. The inductive stability margin is particularly useful as a first measure of vertical controllability as it is independent of the resistivities of the active and passive conducting structures. It is only a function of the mutual inductances between all the conducting structures, vertical derivatives of the mutual inductance between the plasma and each conducting structure, and the currents in the plasma and active coils.

Even with a sufficiently high inductive stability margin and slowed vertical growth due to interactions with surrounding conductors, it's crucial to ensure that the growth rate is slow enough to be actively controlled. This is measured by the normalized growth rate, $\gamma\tau$, where $\gamma$ is the vertical mode's growth rate and $\tau$ is the effective $L/R$ time of the surrounding conducting structures. For controllability, we require $\gamma\tau < 10$. Unlike $m_s$, $\gamma\tau$ depends on the resistances of the surrounding conductors and can be computed using either a massless approximation or with-mass corrections. FUSE employs the massless stability model originally developed in TokSys, as with-mass corrections are generally significant only near or beyond the threshold for ideal instability. The constraints on the metrics are chosen to be well away from this threshold, making the massless stability model sufficient for machine design and optimization.

Although there's a vision to eventually integrate first principles MHD stability codes into FUSE, or even develop them in Julia for optimal compatibility and performance, the immediate emphasis is on enhancing, documenting, and structuring the semi-empirical models into a comprehensive library that correlates models with specific plasma states. In certain situations, semi-empirical models remain at this time the sole available option. Their rapid computation speed makes them ideal for swift prototyping and the development of control strategies to prevent instabilities. However, employing higher-fidelity MHD stability codes will be crucial for finalizing FUSE solutions.

%----------
\subsection{Heating and Current Drive}
\label{models:hcd}
%----------
The mechanisms employed to heat the plasma and drive the current are crucial to achieve and maintain the fusion conditions. FUSE offers analytical models tailored to mock the heating, current, particle, and momentum profiles of various auxiliary systems, including electron (EC) and ion (IC) cyclotron radio frequency heating, lower hybrid (LH) microwaves, neutral beams (NB) and pellet injection.

The analytical models in FUSE use simple Gaussian profiles with user-defined deposition radius and width. The profiles are normalized such that the total power deposited in the plasma matches the launched power, accounting for transmission and generation losses. The current drive efficiently for the different current drive schemes is based on the parametrizations defined in Ref.~\cite{Stambaugh2011}.

In addition to these analytical models, FUSE has been loosely coupled to the RABBIT \cite{weiland2018} neutral beam code, which is real-time capable (taking $\sim$25 ms per time step) and extensively benchmarked with the more sophisticated NUBEAM \cite{pankin2004} code. Importantly, the use of RABBIT allows for the accurate modeling of the time-dependent build-up of the fast-ion distribution, which occurs on the timescale of the characteristic slowing-down time of fast neutral particles as they interact with the bulk plasma.

Pellet modeling is also available as analytic model in FUSE, parametrized as a skewed $\beta$ distribution. Additionally, development of a pure Julia implementation of the Pellet Ablation Module \cite{mcclenaghan2023} is underway. The initial implementation of the {\tt PAM.jl} package will be tailored to particle fueling, as opposed to disruption mitigation.

%----------
\subsection{Power plant}
\label{models:bop}
%----------
FUSE handles the sizing of radial build on both the high and low magnetic field sides of the power plant. Specialized models are available to monitor and address stresses on the power plant's structures or components, ensuring their safety and longevity. Additionally, the power plant's magnetic poloidal coil systems, responsible for plasma shaping, has dedicated optimization workflows. The power plant's structural components are accounted for, as are models dedicated to the distribution and behavior of neutrons produced during fusion. The breeding blanket functions of absorbing fusion neutrons and tritium breeding are also modeled, along with the first wall and divertor components that handle the heat and particles exhaust from the plasma.

%=============
\subsubsection{Radial and cross sectional build}
\begin{figure}
\includegraphics[width=\textwidth]{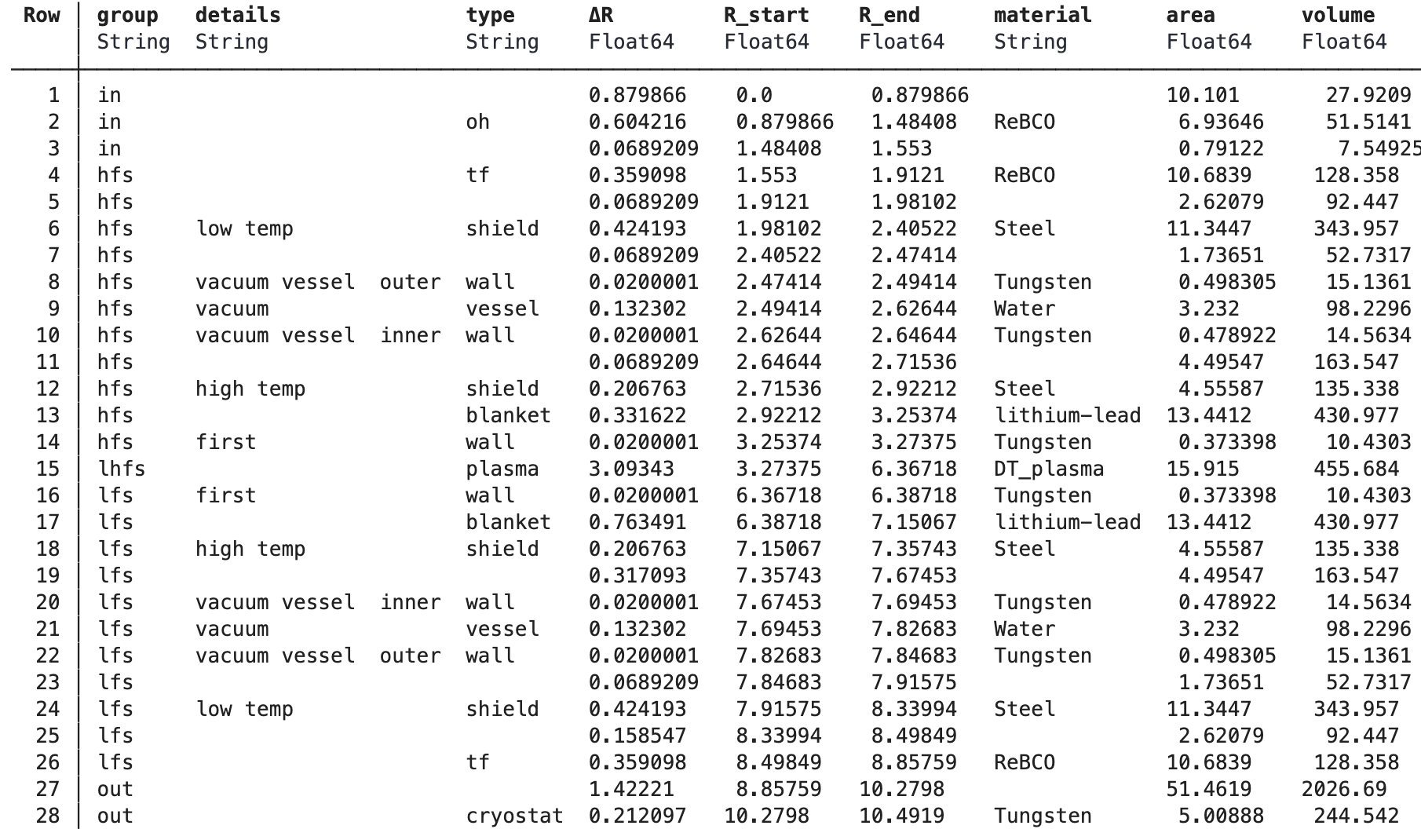}
\caption{Radial build of a prototypical FPP power plant in FUSE, showing each of the layers' properties including  their high and low field side thicknesses and materials.}\label{fig:layers_table}
\end{figure}

%=============
FUSE's cross-sectional build of the power plant is composed of an arbitrary number of concentric layers that surround the plasma (the layers outside the toroidal field coils, like the Ohmic solenoid or the cryostat being the exceptions), as summarized for example in Fig.~\ref{fig:layers_table}. To each of these layers is assigned a high-field-side and low-field-side thickness, which can be different and individually adjusted to satisfy different engineering requirements (eg. neutronics, mechanical stresses, strength of the magnetic field, superconductor currents, pulse duration, ripple, costs, etc.).

\begin{figure}
\includegraphics[width=0.75\textwidth]{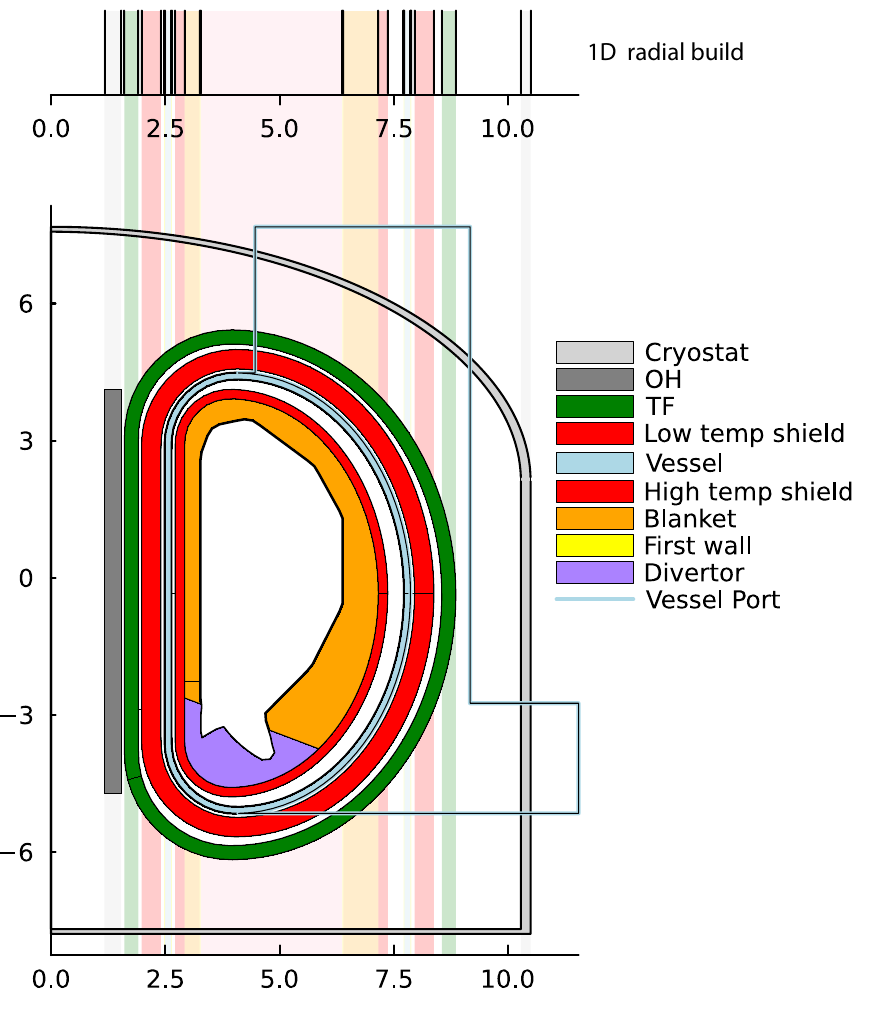}
\caption{Radial and cross-sectional build from the central solenoid to the cryostat of a prototypical FPP power plant in FUSE, illustrating the radial build extending to the cross-sectional build and showcasing the concentric shapes from the plasma to the TF coils, including the added divertor region.}\label{fig:build}
\end{figure}

Going from the radial build to the 2D cross section as can be seen in Figure \ref{fig:build} FUSE supports a wide range of shaping options for the layers' cross-sectional outlines. While each of the layers' outlines could in principle have a parametrization of arbitrary complexity, for most of the exploratory FPP studies it is sufficient to use a simple parametrization of the toroidal field coils shape, and the outline of the first wall is either input externally, or is automatically built around the plasma equilibrium at some reference time-slice. The layers between these the first-wall and the TF coils are designed to be convex, and maintain a constant distance from the adjacent layers. This solution allows for rapid and robust creation of realistic cross-sectional builds, using only a minimal set of free parameters.

In addition, FUSE defines some structures that further partition these concentric layers in the poloidal plane. In particular, the divertor structures are assigned based on the saddle and strike points of a reference equilibrium. FUSE supports both single as well as double divertors. Divertor structures also limit the poloidal extent of the breeding blanket (see Fig.~\ref{fig:layers_table}).

Finally, FUSE can define the extent of both horizontal and vertical ports to be used for maintenance. The sizing of these ports is calculated to allow for sections of the blanket to pass between the TF coils and the port openings. Similarly for the divertor maintenance ports.

\begin{figure}
\includegraphics[width=\textwidth]{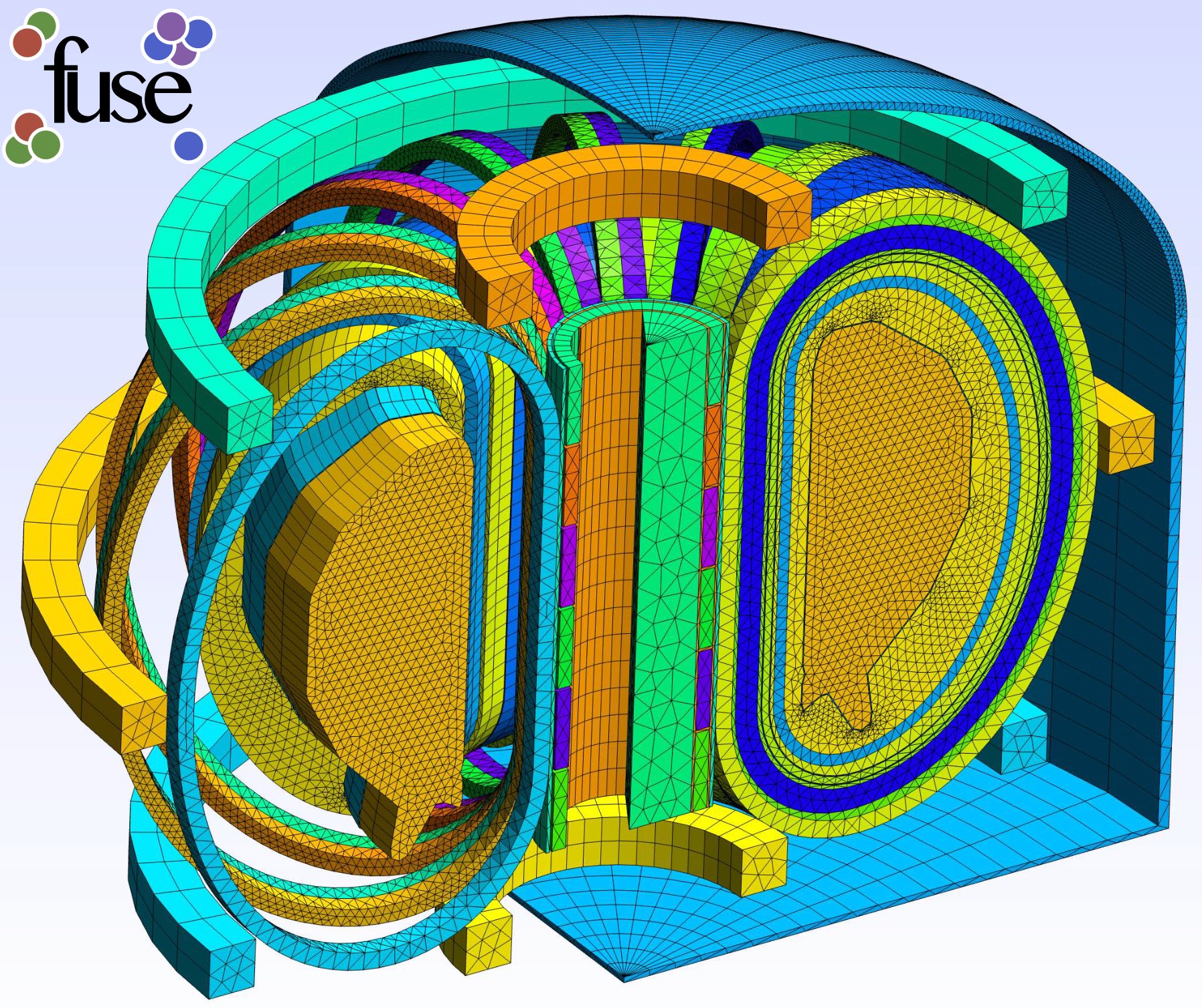}
\caption{CAD and resulting 3D mesh of a FUSE 2D build, generated via the Gmsh Julia API.}\label{fig:FUSE_3D}
\end{figure}

From the 2D build, FUSE can generate a 3D CAD and mesh via Gmsh Julia API (see \ref{fig:FUSE_3D}). Gmsh is an open source 3D finite element mesh generator with a built-in CAD engine and post-processor. At the time of writing this capability does not have an application yet, but in the future, we expect the resulting 3D mesh to be used to execute finite element analysis and other high fidelity simulations. In the future, a translation from the Gmsh internal mesh representation to the IMAS General Grid Descriptor (GGD) model will be needed to store the information in the IMAS data dictionary.

%=============
\subsubsection{Solid mechanics}
\begin{figure}
\includegraphics[width=\textwidth]{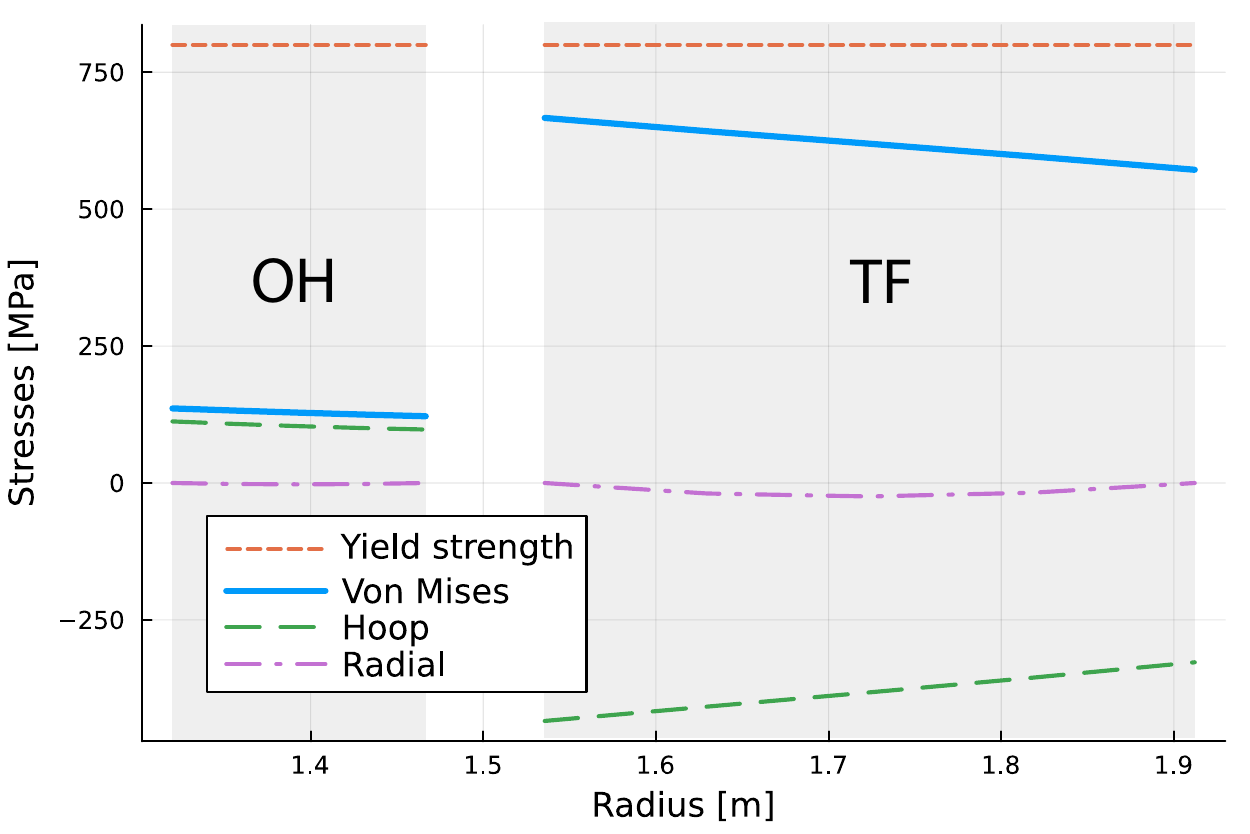}
\caption{Mechanical stress assessment of the TF and OH coils. }\label{fig:stresses.pdf}
\end{figure}

%=============
The central stack is a crucial focal point due to the interplay of magnetic forces originating from the Ohmic Heating (OH) and Toroidal Field (TF) coils. The OH coil is subjected to substantial magnetic pressures acting radially outward, while the TF coils induce strong compressive and tensile stresses within the stack as illustrated in Figure~\ref{fig:stresses.pdf}. The combination of the forces from these coils makes the central stack a critical area of focus in any preliminary mechanical stress assessment of a tokamak.

FUSE solves a 1D solid mechanics linear problem to evaluate the mechanical stresses on the center stack of the device and optimize the toroidal field (TF) coils and Ohmic (OH) solenoid magnets for their size as well as their ratio of steel to (super)conductors material. This model supports different center stack configurations, with a center plug and having the option to buck the TF and OH against one another. FUSE's 1D solid mechanics model has been compared against full 3D simulations of center stacks run in GATM \cite{leuer2023}, showing excellent agreement.

An interface has been developed to streamline the process of information transfer from FUSE to GATM (General Atomics Tokamak Model), a finite element model based on COMSOL that employs parametric CAD generation for assessing the performance of tokamak magnet-plasma systems. GATM consists a 2-D axisymmetric module characterizing magnetic properties of the poloidal fields system, including the plasma representation. The 2D module contains a central solenoid structural model for stress and deflection assessment. The toroidal field coil is characterized by a 3D magnetics module for static magnetic field calculations and simulation of non-axisymmetric ripple and error fields. The magnetics models provide input to the 3D TF structural module, which simulates both in-plane and out-of-plane stress. A 3-D center post module uses input from the other three modules to evaluate performance of a superconducting coil including details of case, winding-pack and superconducting cable. FUSE's 1D solid mechanics model has been compared against GATM simulations, showing excellent agreement.

%=============
\subsubsection{Poloidal field coils optimization}
\label{ActorPFcoilsOpt}
%=============
Optimization of the poloidal field (PF) coils placement can be notoriously time consuming, as it involves updating the Green's functions tables (typically 100,000+ mutual inductance calculations) and solving a free-boundary equilibrium problem for each PF coil configuration tested by the optimization algorithm. Instead, FUSE adopts an alternative approach, designed to minimize the least-square error between the plasma and shaping coils contribution to the poloidal flux evaluated at the last-closed flux surface. This method is particularly efficient since the plasma contribution is held constant and evaluated with a fixed-boundary equilibrium solver, while the contributions to the coils in the vacuum region needs to be evaluated only on the boundary of the fixed equilibrium solver (typically ~ 1000 mutual inductance calculations) for each PF coil configuration. 

\begin{figure}[htp]
    \centering
    % First image
    \begin{minipage}[b]{0.45\textwidth}
        \includegraphics[width=\textwidth]{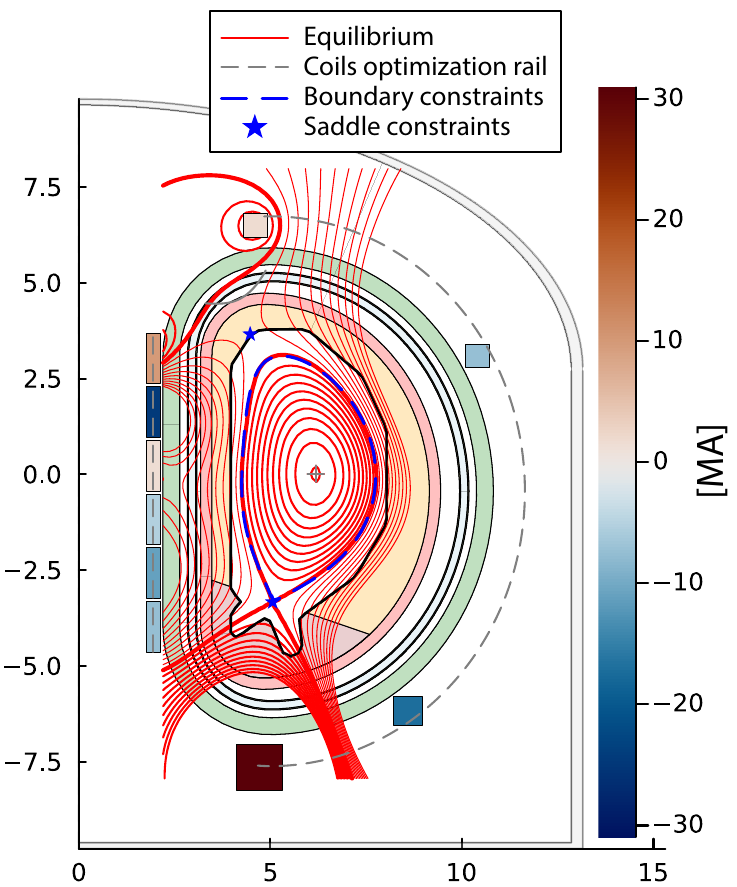}
    \end{minipage}
    \hfill % Space between the images
    % Second image
    \begin{minipage}[b]{0.45\textwidth}
        \includegraphics[width=\textwidth]{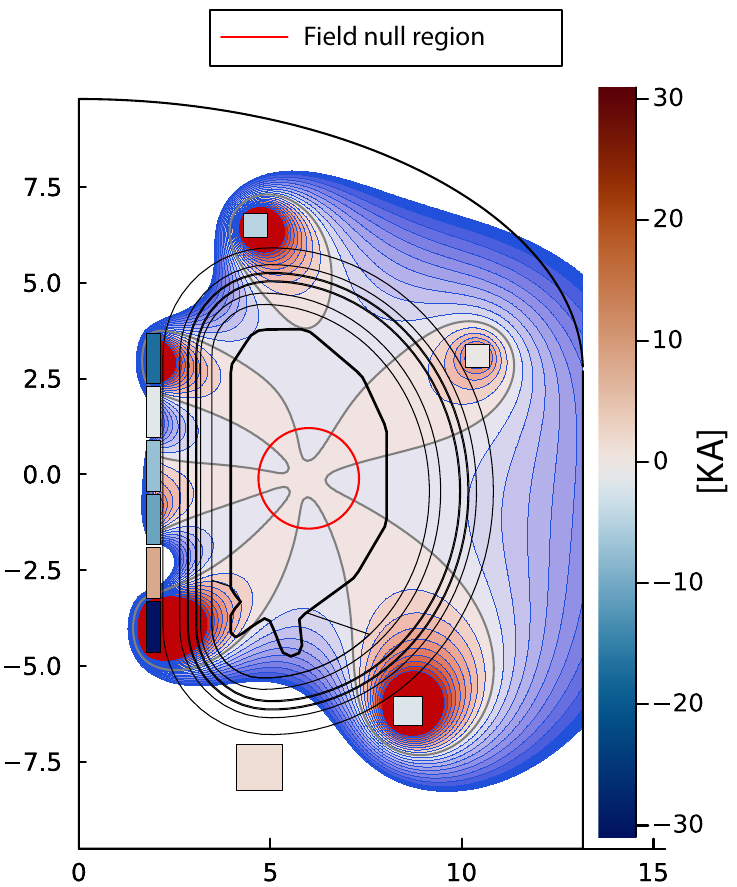}
    \end{minipage}
    \caption{Results of a PF coil optimization in FUSE. The left panel show the initial closed boundary equilibrium solution extended to vacuum with single lower null is shown in blue, while the red contours are the solution that is achievable with the given PF coils configuration. The bottom panel, shows the target poloidal field and the target null region (red circle) in the pre-magnetization stage. The optimized solution is found in 250ms starting from uniformly distributed coils.} % Main caption for both images
    \label{fig:pf_coils_opt}
\end{figure}

The PF coil optimization in FUSE considers the maximum currents that the PF coils can carry, supports the addition of iso-flux, strike-points, and saddle points (ie. X-points) constraints. The optimization can minimize the total error with respect to multiple target equilibria, including the pre-magnetization field null region, results of an example optimization can be seen in Fig.~\ref{fig:pf_coils_opt}. In FUSE, the PF coils are positioned along guiding rails that follow the shape of inner layer where the coils are defined. These guiding rails can have breaks or gaps, specifically designed to designate areas where PF coils cannot be placed due to constraints like port access or maintenance requirements (at the time of writing modeling of maintenance requirements is planned but not yet implemented in FUSE).

%=============
\subsubsection{Neutronics and blanket}
%=============
FUSE offers the ability to perform neutronics calculations with a tightly coupled reduced model, as well as via connection to external 3D Monte Carlo codes. In both cases, FUSE provides a realistic volumetric neutron source consistently calculated from plasma core profiles and equilibrium for both D-T and D-D fusion reactions.

\begin{figure}[ht]
\includegraphics[width=\textwidth]{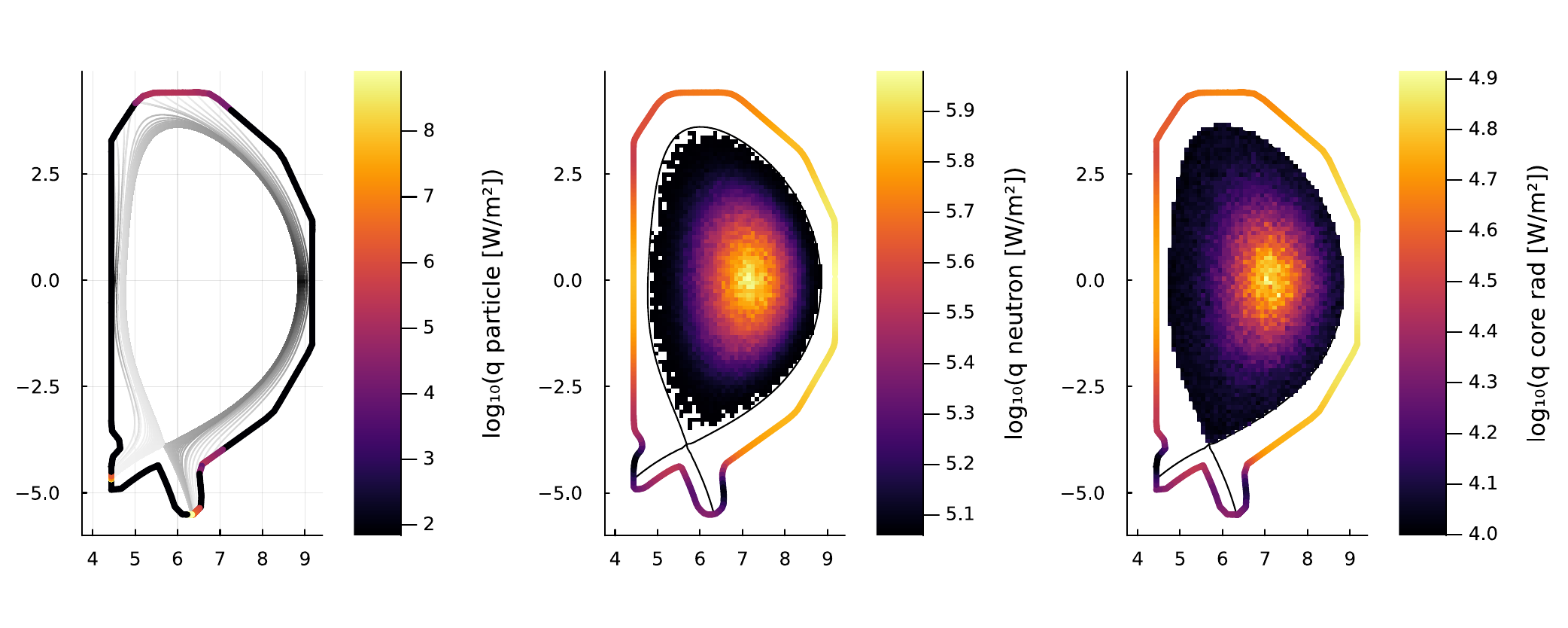}
\caption{Example of resulting plasma, neutron and radiative heat flux on the divertor and first walls calculated by FUSE}\label{fig:loads_cx}
\end{figure}

FUSE's reduced neutronics model traces the direct path of neutrons within the vacuum chamber to the first wall, determining the initial neutron wall loading. Figure~\ref{fig:loads_cx} shows an example of a neutron first wall flux result as calculated by FUSE. Within the materials, this inherently 3D neutronics transport problem is approximated as the superimposition of 1D neutronics simulations where the materials' thicknesses seen by each neutron changes depending on the machine's poloidal geometry and the neutron angle of incidence. To further expedite the 1D simulations, a surrogate model, constructed from a database of 1D Monte Carlo neutronics simulations, is employed. The tritium breeding ratio (TBR) derived from this condensed model match within a 15\% margin 3D Monte Carlo simulations that use the same geometry and materials. FUSE leverages this reduced model to efficiently optimize the thickness of the first wall, breeding blanket, and neutrons shield. It also fine-tunes the $^6$Li/$^7$Li enrichment to meet a specified TBR while minimizing the neutron flux that escapes the blanket.

At the high-fidelity end of the modeling spectrum, FUSE can export its geometric representation of the device and neutron source to setup a full OpenMC \cite{romano2015} 3D Monte Carlo simulation. This workflow relies on the Paramak package to 1) generate the 3D geometry mainly by revolving the cross-sectional outlines of the FUSE's internal geometric representation of the device; 2) convert such 3D geometry to a OpenMC-compatible input via DAGMC; 3) execute the OpenMC simulation and postprocess the Monte Carlo simulation tallies. Because of the computational cost of this workflow, its execution is generally triggered manually, on a case by case basis, and the simulation results are not feed back directly into FUSE simulations, but rather used as a high-fidelity verification of the lower fidelity models.

% Bridging these two modeling extremes, FUSE is advancing the development of a deterministic 2D axisymmetric neutronic calculation in pure Julia. While this capability is still in progress, initial results indicate a promising balance between model accuracy and computational speed in cylindrical (not yet toroidal) geometry.

%=============
\subsubsection{Scrape off layer and divertors}
%=============
FUSE can analyze the open field lines of an equilibrium and categorize them as belonging to the near or far SOL, based on they relative distance between the last closed flux surface, the primary x-point, and any subsequent x-points. This information, in addition to knowledge of the strike point locations and angles allows for easy implementation of reduced SOL models, with parameters that are grounded in realistic equilibrium and first wall geometries. This information can then be used to properly size the divertors volumes.

At the lowest order, machine designs can be constrained by global parameters, widely accepted to be representative of the SOL and divertor challenge. A common approach is to limit those global parameters to be within the range of the ITER design (eg. $P_{\rm sep}/R < 15$ MW/m.

Next, is the Lengyel reduced model \cite{lengyel1981}, a simple approximation with some degree of sophistication regarding the radiative characteristics of the atomic species present in the plasma. Lengyel relates the divertor conditions (in terms of the impurity concentration in the divertor, the upstream SOL electron density, and the upstream parallel energy flux density) to the power dissipated. If detachment onset is assumed to occur at the point where plasma power dissipation is complete, then the Lengyel model can be used to calculate the divertor conditions required for detachment onset.

% Add about the stangeby/nichols 2pt model

%----------
\subsection{Facility}
\label{models:facility}
%----------
The operation of a fusion power plant requires more than just the core reactor; it involves a range of supporting systems that are essential to its overall functionality. The FUSE framework provides detailed models to simulate these auxiliary systems, integrating crucial aspects such as heat transfer, cooling systems, and power distribution into the design and optimization process. These models ensure the plant operates efficiently and safely by addressing all necessary subsystems. In addition to technical considerations, FUSE includes comprehensive costing models that estimate the expenses related to construction, operation, maintenance, and decommissioning. By providing insights into the economic aspects of fusion power plants, FUSE ensures that designs are both technically sound and economically viable.

%=============
\subsubsection{Balance of plant}
%=============
FUSE's balance of plant (BoP) modeling is powered by the newly developed {\tt ThermalSystemsModels.jl} package, which is capable of modeling both single-phase and multi-phase fluids, including liquid and gaseous systems. This package defines a range of components, such as heat exchangers, turbines, pumps, compressors, and condensers, which can be configured in a network to simulate Rankine and Brayton cycles of varying complexity. While the {\tt ThermalSystemsModels.jl} package can be used independently and applied to a variety of industries beyond fusion, within FUSE, it is specifically tailored to capture and optimize the power generated from all heat sources in a fusion plant, including the blanket breeder, first wall, and divertors. The solver supports both time-dependent and stationary solutions, allowing for detailed analysis and optimization. An integrated optimizer enables users to determine the optimal coolant flow rates, maximizing heat-to-electricity conversion efficiency. The results produced by {\tt ThermalSystemsModels.jl} have been verified against Thermoflow simulations, ensuring their accuracy and reliability.

ThermalSystemsModels versatility extends beyond its current capabilities. It is designed to be easily expandable, with plans for future incorporation of additional non-fluid components, specifically for mechanical power transmission, generators, and a wide spectrum of electrical systems within the power plant. This comprehensive framework, in theory, has the potential to simulate dynamic models for all energy-consuming and energy-producing subsystems within the fusion power plant.

\begin{figure}[!ht]
{\bf A)}\\
\includegraphics[width=\textwidth]{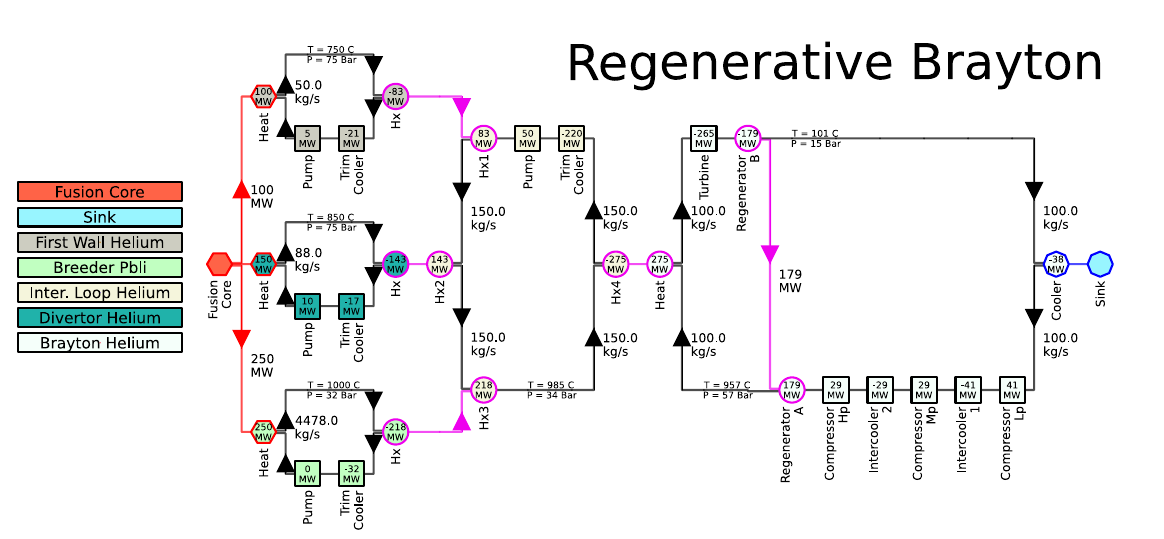}\\
{\bf B)}\\
\includegraphics[width=\textwidth]{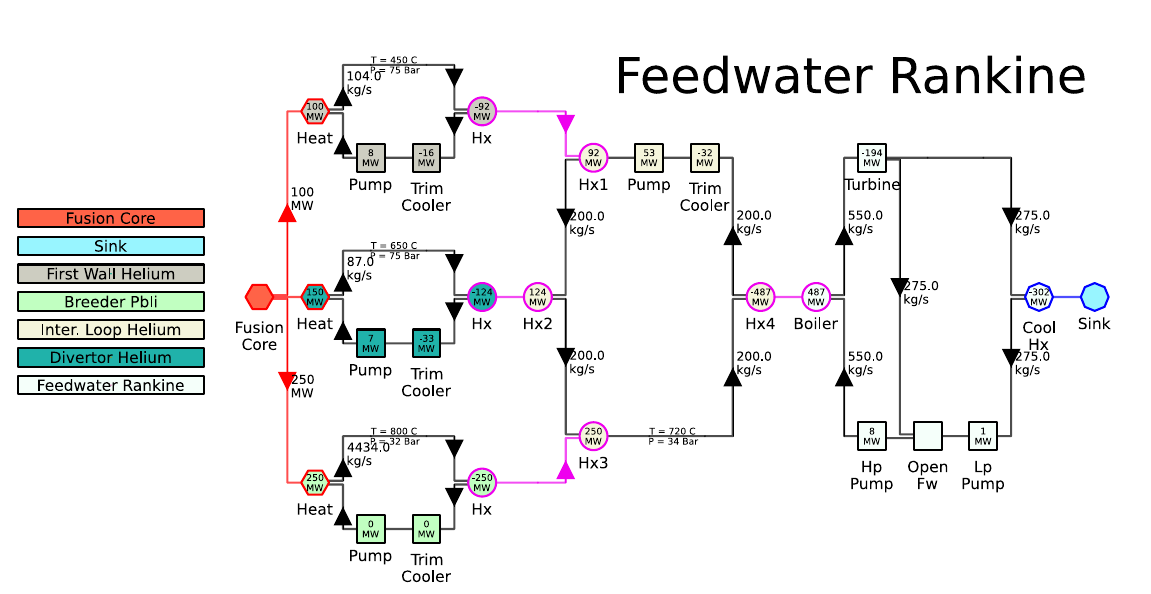}
\caption{Examples of different balance of plants configurations simulated in FUSE: A) depicts a regenerative Brayton Cycle with its distinct components such as heat exchangers and compressors. B) illustrates a Rankine Cycle, emphasizing its feedwater system pathways and associated components. Both systems accept heat from blanket breeder, first wall, and divertors. Together, these diagrams provide a comparative view of the different operational pathways and components inherent to each cycle, underscoring FUSE's versatility in modeling optimal heat-to-electricity conversion.}
\label{fig:RegenBrayton_output}\label{fig:FeedwaterRankine}
\end{figure}

%\subsubsection{Tritium cycle}? % Dave

%=============
\subsubsection{Costing}
\begin{figure}[ht]
A)\\
\includegraphics[width=\textwidth]{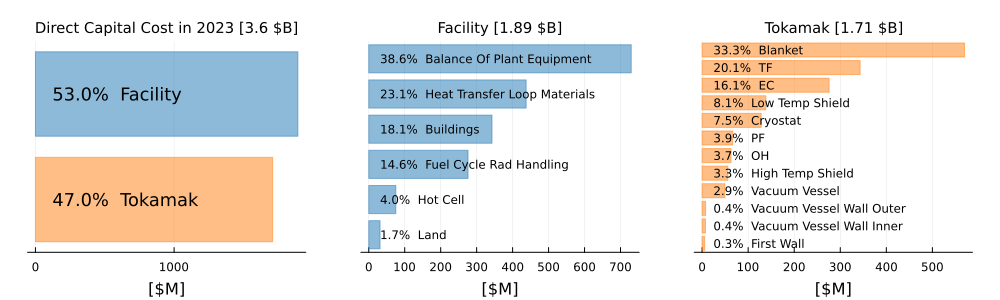}
B)\\
\includegraphics[width=\textwidth]{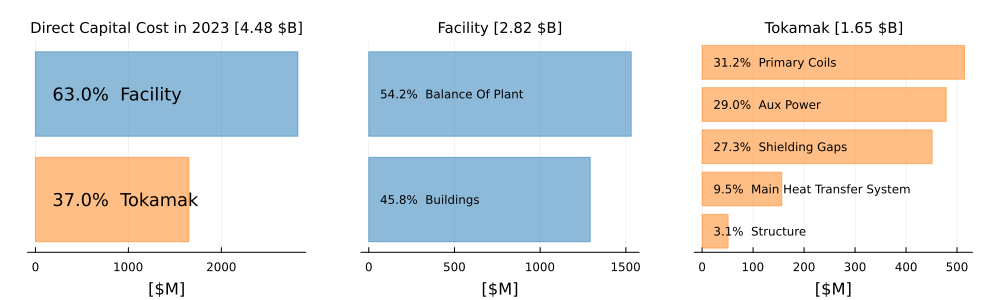}
\caption{Direct capital cost for the GA FPP version 1 \cite{GAFPP} design as calculated by the A) ARIES and B) Sheffield costing models.}
\label{fig:cost_fpp}
\end{figure}

%=============
FUSE contains two distinct models for costing a final plant design, namely the ARIES \cite{waganer2013} and Sheffield \cite{sheffield1986, sheffield2016} models. Each model is individually capable of calculating the direct capital cost of a design as well as its associated levelized cost of electricity. The capital cost includes the construction, operation, maintenance and decomissioning costs. Resulting direct capital costs from the ARIES and Sheffield models for a given FPP design are shown in Figs.~\ref{fig:cost_fpp} respectively.

ARIES was published in 2013, while Sheffield was originally published in 1986 but significantly updated in 2016 following benchmarking against ITER costs. Broadly, the ARIES model is a more modular approach than the Sheffield model, breaking down tokamak and facility costs into smaller groups with more customized costing algorithms. Sheffield tends to predict higher direct capital costs for the same plant design than ARIES. 

Beyond the base models described in Refs.~\cite{waganer2013, sheffield1986, sheffield2016}, FUSE also includes an inflation adjustment calculator. The calculator uses a database of historical inflation rates to normalize the costs returned by each model to their respective costs in present-day dollars. Optionally, costs can also be inflated to a user-determined future construction start year. 

%=======
\section{Stationary and Time-Dependent Modeling}
\label{sec:timedep}
%=======

One of the most prominent features of FUSE lies in its capability to run both stationary and time-dependent simulations. While stationary simulations provide valuable insights into stable operational points, time-dependent simulations establish how to access the conditions identified by stationary simulations, making FUSE a valuable tool for pulse design in both new and existing machines.

The FUSE stationary plasma actor iterates between equilibrium, steady-state current solver, sources, and core transport actors to find a self-consistent core-plasma solution valid as $t \rightarrow \infty$, similarly to the approach in Refs.~\cite{meneghini2017, Meneghini2020}. These components are iterated until a measure of convergence, defined as the relative change in the current and pressure profiles, falls below a given threshold.

The FUSE dynamic plasma actor leverages the natural hierarchy of timescales present in typical tokamak plasmas. The equilibrium (Alfv\'enic) timescale is the fastest, typically on the order of a microsecond, followed by the transport timescale, which ranges from tens of milliseconds to seconds. The longest is the the resistive current-diffusion timescale, particularly in high-temperature burning plasmas, and can range from seconds to minutes. This ordering can be expressed as:

\[
\tau_{\text{equilibrium}} \ll \tau_{\text{transport}} \ll \tau_{\text{resistive}}.
\]

Given this ordering, the QED current diffusion solver advances the safety-factor profile in time, while assuming that equilibrium, sources, density, rotation, and temperature evolve adiabatically, effectively reaching a steady state between each QED solution. Under these assumptions, transport continues to be treated as a flux-matching problem, as in the stationary plasma actor. 

To accurately model transport dynamics that occur on timescales comparable to the confinement time—such as L/H transitions or the dynamic formation of internal transport barriers—FUSE uses a 1st-order, backward-Euler implicit time advance, including the time derivative terms of the transport equations \cite{sugama1998} as an additional source term in the flux-matching solver. The separation of timescales, combined with the implicit treatment of the transport time derivatives, permits solving the transport problem over time steps that are greater than $\tau_{\text{transport}}$, thus greatly reducing the overall computational cost of performing time-dependent simulations.

Sources generally interact with the plasma on a faster timescale than the confinement time. The thermalization of fast ions, such as those from neutral beam injection (NBI), ion cyclotron (IC) heating, or fusion-produced alpha particles, occurs on a timescale that is close to, but still faster than, the confinement time in typical reactor-sized machines. In an ITER baseline case, for example, the energy confinement time is approximately 2.5 seconds, while the thermalization time of alphas is around 0.5 seconds. Since FUSE currently neglects the lag between the birth of alphas and their thermalization, this imposes a lower bound on the simulation time step, which should be set greater than their thermalization time.

Figure~\ref{fig:ITER_time_dep} illustrates a time-dependent simulation of an ITER plasma, ranging from 50 to 350 seconds. Between 150 and 250 seconds the current is ramped from 13MA to 15MA via a PID controller that operates on the loop voltage boundary condition provided to QED (Fig.~\ref{fig:ITER_time_dep}[1,1]). Simultaneously, all auxiliary systems are also ramped from half to full power, as reflected in the changes in current, heating, and particle sources (Fig.~\ref{fig:ITER_time_dep}[2,1:4]). The transport calculation ensures that the transport fluxes remain consistent with the total sources, maintaining flux matching (see Fig.~\ref{fig:ITER_time_dep}[3,2:4]). The temperature and density profiles increase in response to the higher auxiliary injections (see Fig.~\ref{fig:ITER_time_dep}[1,3:4]). The evolution of the safety factor, which began in a fully relaxed state, changes rapidly as the current increases, and more slowly after the ramp ends, allowing the ohmic current to penetrate the plasma (Fig.~\ref{fig:ITER_time_dep}[3,1]). In this simulation, the plasma shape is held fixed, yet a small change in the Shafranov shift can be observed in the equilibrium as the pressure increases. Additionally, the PF coil currents increase alongside the plasma current (see Fig.~\ref{fig:ITER_time_dep}[1,2]). Subplots in Figure~\ref{fig:ITER_time_dep} are indicated using the [row,column] format, where `:` is used to indicate a range, such as [2,1:4] representing subplots in the second row, from the first to the fourth column.

\begin{figure}[ht]
    \animategraphics[loop,autoplay,poster=60,width=\textwidth]{12}{figures/ITER_time_dep/frame}{000}{120}
\caption{Animation, showing an example of a time-dependent simulation carried out in FUSE for a combined current and injected power ramp on ITER. Please note that the embedded animation is supported in PDF readers like Adobe Acrobat Reader (version 9.0 or later) and PDF-XChange Viewer. Other PDF readers may not display the animation.}\label{fig:ITER_time_dep}
\end{figure}

%Time-dependent simulations facilitate the in-depth development and testing of control systems, enabling engineers to devise, validate, and refine control strategies.

% Ongoing work aims at:
% \begin{itemize}
% \item handling L-H transition, particularly relating to the handling of the pedestal
% \item operate on power supplies instead of directly on coil currents
% \item include vessel response
% \item Free boundary equilibrium solver
% \item Grad Hogan formulation
% \item connection to TokSys controllers (fxp)
% \end{itemize}

At the moment, FUSE time-dependent simulations use TEQUILA fixed-boundary equilibrium calculations along with the Green's function method to determine the necessary PF coil currents. This requires that the equilibrium shape be specified (potentially as a function of time) and lacks inductive dynamics in the PF coils and surrounding conducting structures. A more realistic model is currently underdevelopment that would make use of a new Julia-based free-boundary equilibrium solver, together with time-dependent transport, inductive coupling throughout the whole device, and actuator control. In essence, this will create a Grad-Hogan-like solver combining the physical accuracy and computational efficiency of FUSE's time-implicit, flux-matching transport solver with sophisticated feedback control through coupling to GA's TokSys. This will allow for high-fidelity predictive modeling of controllability and scenario access, and will form the foundation for future pulse-design and digital twin research with FUSE.

%=======
%=======
\section{FUSE modeling and design studies}
%=======
%=======
FUSE \emph{studies} perform self-contained studies, typically encompassing multiple simulations. Currently supported studies include validation and constrainted multi-objective optimization.

% figures comparing simple and higher fidelity solutions for the same case?

%----------
\subsection{Plasma validation study}
\begin{figure}[ht]
    \centering
    \includegraphics[width=\textwidth]{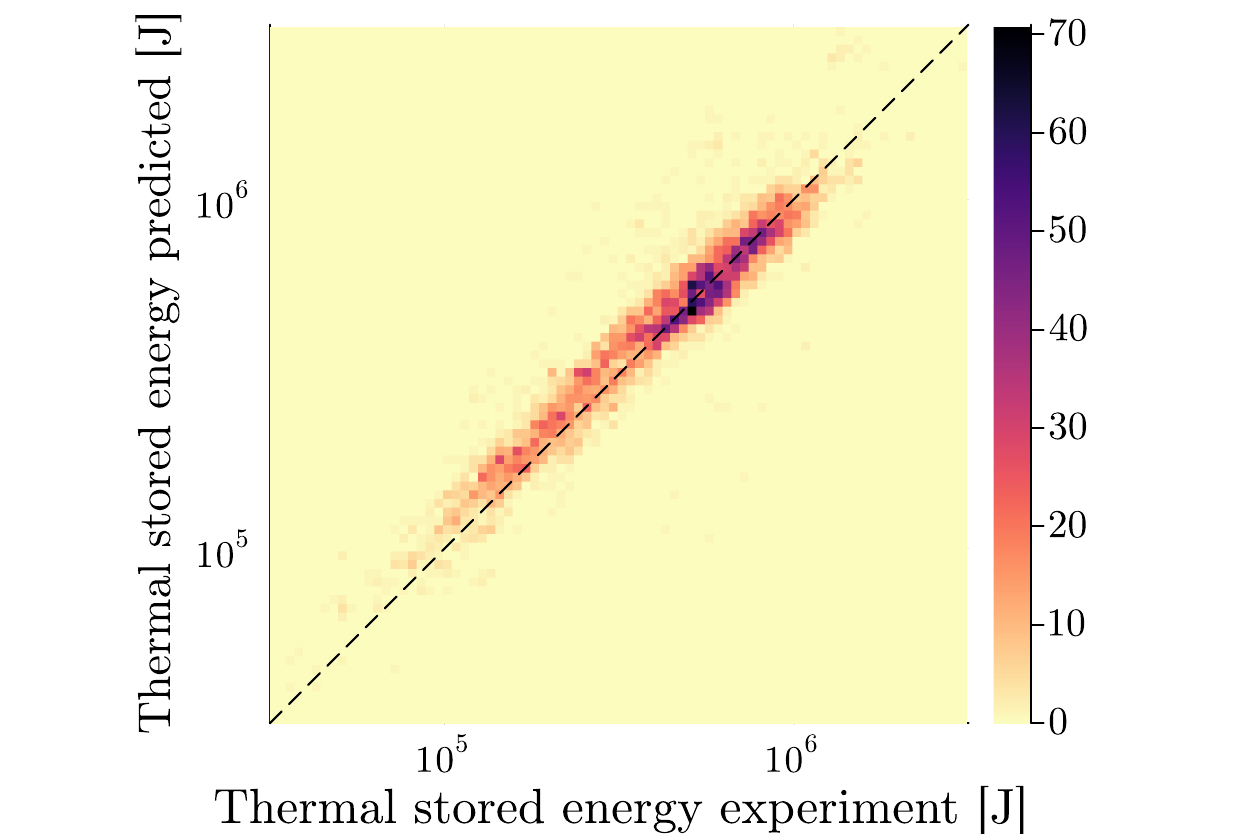}
\caption{Experimental thermal stored energy compared to the predicted thermal stored energy using the FUSE flux-matching transport solver on a DIII-D database of N = 6571 time-slices, this is using the TGLF-NN SAT 1 EM model on a 9 points equally spaced $\rho$ grid between 0.1 and 0.9, the mean relative error of the experimental to predicted stored energy is 15\%.}\label{fig:Flux_matcher_exp}
\end{figure}

%----------
To extend the validation of the FUSE modeling beyond individual use cases, the plasma models withing FUSE have been applied to entire datasets. In particular, to validate the flux-matching transport modeling abilities within FUSE, we examined an experimental 2019 database of over 6500 DIII-D time-slices \cite{neiser2020big}. This database encompasses experimental heating, current drive sources, equilibrium, and fitted kinetic profiles. The thermal energy predicted by the FUSE flux-matching transport solver is in quantitative agreement with the experimental thermal stored energy (see Fig. ~\ref{fig:Flux_matcher_exp}). The predictions exhibit a mean relative error of roughly 15\%, with the most significant data cluster fitting within a 10\% range. Notably, the turbulent fluxes were modeled using the re-trained TGLF-NN SAT1 including electromagnetic contributions~\cite{Neiser22}, which is a fast and accurate neural network representation of the quasi-linear transport model TGLF \cite{Staebler07, Staebler16, Staebler21}.% For context on processing speed, the 6571-run database required approximately 270 CPU hours on Intel(R) Xeon(R) CPU E5-2650 hardware.

Furthermore, the study initially conducted using the STEP modeling in Ref.~\cite{Slendebroek2023} was extended to FUSE: 600 discharges from the ITPA global H-mode database were analyzed, employing the same filtering and processing methods detailed in the original study. These methods ensured that the data used was aligned with regions where plasma models are expected to be valid. Comparison of our FUSE results with the experimental energy confinement time revealed a mean relative error similar to that found with the STEP results. Notably, certain discrepancies were observed in the predictions for small machines, emphasizing the need to further explore and understand the plasma behaviors in different confinement regimes. This renewed investigation not only strengthens our confidence in FUSE to accurately predict fusion power outputs for larger devices but also reinforces the importance of refining plasma models for enhanced accuracy.

% TIM: add picture with results

%----------
\subsection{Optimization of fusion power plant design}
\label{sec:multi_objective_optimization}
%----------
One of the key applications of FUSE has been its use for whole facility optimization of a fusion power plant. The design of fusion power plants is inherently complex, characterized by the intricate interplay of numerous components and subsystems. Stakeholder interests are diverse, with engineers, investors, and policymakers often prioritizing different objectives. While engineers might emphasize safety and efficiency, investors could lean towards cost-effectiveness and return on investment.

In such a multifaceted environment, the traditional approach of optimizing a single parameter often falls short. Instead, a more holistic perspective is required, one that can balance various competing objectives. This is where multi-objective optimization shines. By allowing for the simultaneous consideration of multiple goals, it ensures that the design process doesn't overly focus on one aspect at the expense of others. One of the standout features of multi-objective optimization is its ability to provide a Pareto front, a set of optimal solutions where any improvement in one objective would result in a degradation in another. This not only offers a clear picture of the trade-offs between different objectives but also empowers decision-makers to select a solution that best aligns with their priorities. Such a comprehensive approach is especially beneficial in fusion plant design, ensuring that the optimization of one aspect doesn't inadvertently compromise another.

\begin{figure}[!ht] \includegraphics[width=0.49\textwidth]{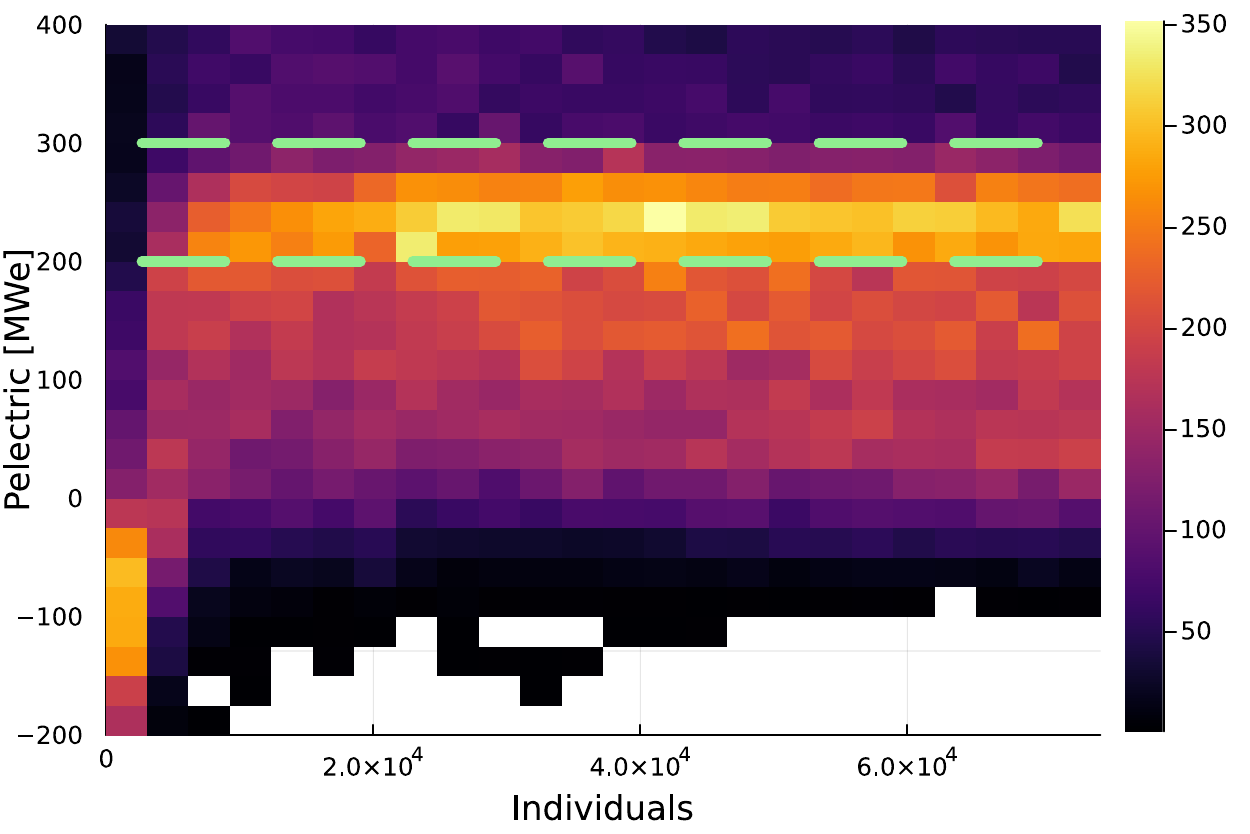}\hfill \includegraphics[width=0.49\textwidth]{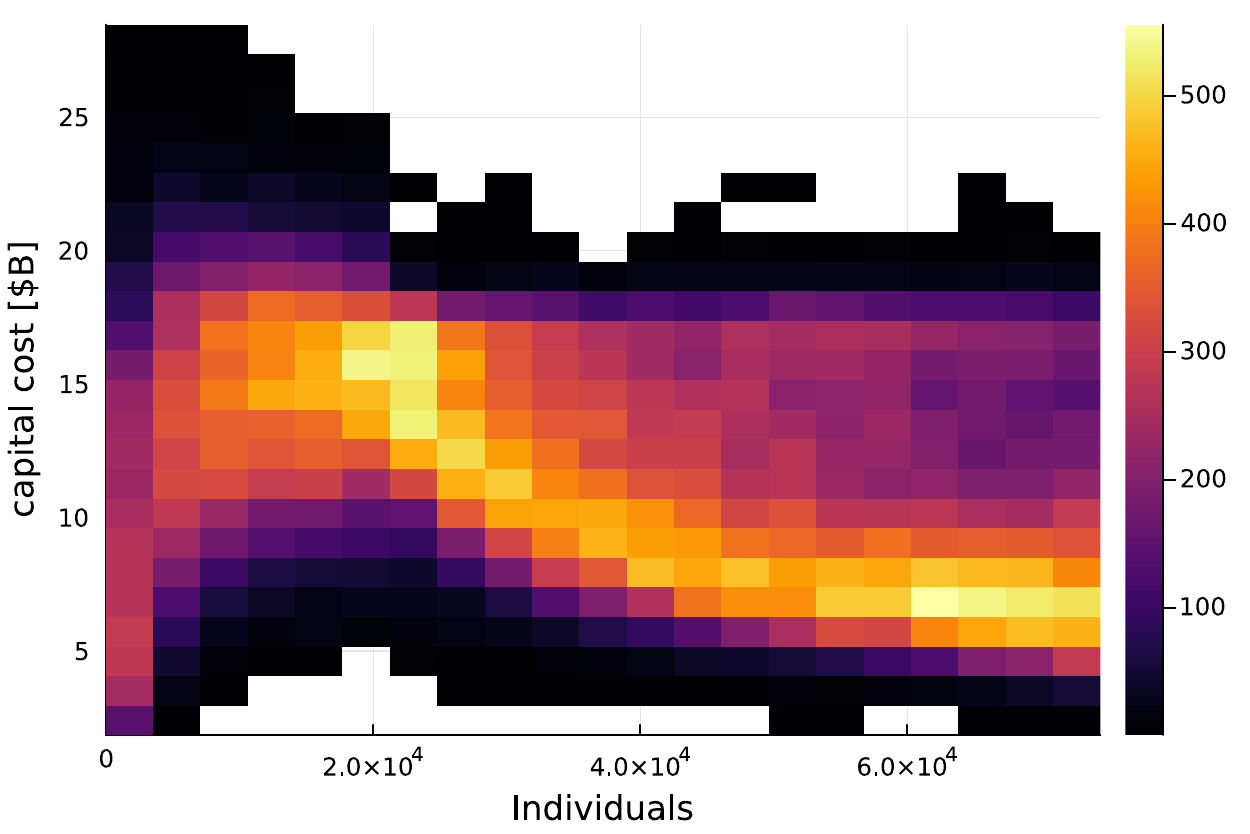} \caption{2D histograms illustrating the progression of individuals across generations within the genetic algorithm of the multi-objective constrained optimization process. The left plot demonstrates the initial struggle to satisfy the $P_{\text{electric}}$ constraint, with the majority of individuals eventually meeting this requirement as the optimization progresses. The right plot tracks the genetic algorithm's drive to find solutions at lower cost. First the optimizer finds high-cost (large-scale) designs that satisfy the $P_{\text{electric}}$ constraint, and later transitioning to more cost-effective designs that continue to adhere to the imposed constraints.} \label{fig:constraints_and_opjectives} \end{figure}

The FUSE multi-objective optimization workflow can execute any actor or user-defined function. Typically, the whole facility actor is used to execute all the models described in Sec.~\ref{sec:actors} and provide a holistic view of a machine design. A comprehensive library of objectives and constraints is available, while also offering the flexibility to define custom specifications. Within the FUSE optimization run, any input parameter can be adjusted. This includes not only floating-point values but also simulation switches, facilitating optimizations that span discrete choices, such as HTS versus LTS, all within a unified simulation framework. FUSE makes use of genetic algorithms implemented in the Julia Metaheuristics.jl package \cite{metaheuristics2022} to evolve a population of machine designs towards the Pareto front, as illustrated in Fig.~\ref{fig:pareto_animation}.
The objectives, constraints, and actuators used in this simulation are summarized in Tbl.~\ref{tab:objectives_constraints_actuators}, while Fig.~\ref{fig:constraints_and_opjectives} demonstrates the evolution of individuals across generations within the genetic algorithm of the multi-objective constrained optimization process.

\begin{table}[!ht]
\centering
\begin{tabular}{ccc}
    \begin{minipage}{0.26\textwidth}
        \centering
        \begin{tabular}{|c|}
        \hline
        \textbf{Objectives} \\ \hline\hline
        Minimize capital cost \\ \hline
        Maximize $q_{95}$ \\ \hline
        \end{tabular}
    \end{minipage}
    &
    \begin{minipage}{0.38\textwidth}
        \centering
        \begin{tabular}{|c|}
        \hline
        \textbf{Constraints} \\ \hline\hline
        $P_{\text{electric}} = 250 \pm 50$ MW \\ \hline
        TBR $= 1.1 \pm 0.1$ \\ \hline
        $P_{\text{sol}} / P_{LH} > 1.1$ \\ \hline
        $P_{\text{sol}} / R < 15$ [MW/m] \\ \hline
        $j_{\text{crit,TF}}$, $j_{\text{crit,OH}}$, $j_{\text{crit,PF}} < j_{\text{crit,material}}$ \\ \hline
        $\sigma_{\text{crit,TF}}$, $\sigma_{\text{crit,OH}}$, $\sigma_{\text{crit,PF}} < \sigma_{\text{material}}$ \\ \hline
        \end{tabular}
    \end{minipage}
    &
    \begin{minipage}{0.25\textwidth}
        \centering
        \begin{tabular}{|c|}
        \hline
        \textbf{Actuators} \\ \hline\hline
        $5.0 < R_0 < 10.0$ [m] \\ \hline
        $3.0 < B_0 < 15.0$ [T] \\ \hline
        $4.0 < I_p < 22$ [MA] \\ \hline
        $1.5 < \kappa < 2.2$ \\ \hline
        $0.0 < \delta < 0.7$ \\ \hline
        $1.1 < Z_{\text{eff, ped}} < 3.5$ \\ \hline
        $0.4 < f_{GW, \text{ped}} < 0.85$ \\ \hline
        \textit{Impurity:} Ne, Ar, Kr \\ \hline
        $0 < P_{EC} < 100$ [MW] \\ \hline
        $0 < \rho_{EC} < 0.9$ \\ \hline
        $0 < P_{NB} < 50$ [MW] \\ \hline
        TF shapes \\ \hline
        \end{tabular}
    \end{minipage}
\end{tabular}
\caption{Objectives, constraints, and actuators for a sample FPP multi-objective constrained optimization}
\label{tab:objectives_constraints_actuators}
\end{table}

\begin{figure}[ht]
\centering
    \animategraphics[loop,autoplay,poster=105,width=0.5\textwidth]{12}{figures/pareto_animation/frame_}{0}{105}
\caption{Animation, showing an example of a design optimization study carried out in FUSE, with the objectives of minimizing capital cost and maximizing $q_{\rm 95}$ (a proxy for risk) with the constraints of fixed electric power generation and flattop duration. The figure illustrates the convergence of the genetic algorithm on the Pareto front. Multi-objective optimization, by its very nature, can seamlessly incorporate different perspectives, leading decision makers to make a more informed decision, and to a design that enjoys broader acceptance and support. Please note that the embedded animation is supported in PDF readers like Adobe Acrobat Reader (version 9.0 or later) and PDF-XChange Viewer. Other PDF readers may not display the animation.}
\label{fig:pareto_animation}
\end{figure}

%----------
\section{Conclusion}
%---------- # needs work
The Fusion Synthesis Engine (FUSE) represents a significant advancement in the integrated design and optimization of fusion power plants. By combining first-principle models, machine learning techniques, and reduced models within a unified, modular framework, FUSE addresses the limitations of traditional design approaches that rely heavily on simplified system codes. The ability to perform both stationary and time-dependent simulations, along with a hierarchical approach to model fidelity, allows for comprehensive exploration of design and operational scenarios. FUSE's open-source development in Julia not only enhances computational efficiency but also fosters collaboration within the fusion research community, ensuring that the framework remains adaptable and continuously evolving.

Through the application of multi-objective optimization, FUSE enables a balanced consideration of various design trade-offs, leading to solutions that are both technically robust and economically viable. The extensive validation against experimental data and the ability to rapidly iterate designs positions FUSE as a crucial tool in the acceleration of fusion power plant development. As the fusion community progresses toward the realization of commercially viable fusion energy, FUSE's capabilities in integrated modeling and optimization will be indispensable in navigating the complex design space of future fusion pilot plants.

%-------
\section*{Acknowledgements}
%-------
Work supported by General Atomics corporate funding.

%-------
\section*{References}
%-------
\printbibliography[heading=none]

@article{Stambaugh2011,
author = {R. D. Stambaugh and V. S. Chan and A. M. Garofalo and M. Sawan and D. A. Humphreys and L. L. Lao and J. A. Leuer and T. W. Petrie and R. Prater and P. B. Snyder and J. P. Smith and C. P. C. Wong},
title = {Fusion Nuclear Science Facility Candidates},
journal = {Fusion Science and Technology},
volume = {59},
number = {2},
pages = {279-307},
year  = {2011},
publisher = {Taylor & Francis},
}

@article{Buttery2022,
  title={The advanced tokamak path to a compact net electric fusion pilot plant},
  author={Buttery, RJ and Park, JM and McClenaghan, JT and Weisberg, DB and Canik, J and Ferron, J and Garofalo, AM and Holcomb, CT and Leuer, JA and Snyder, PB and others},
  journal={Nuclear Fusion},
  volume={62},
  number={12},
  pages={128002},
  year={2022},
  publisher={IOP Publishing}
}

@article{Meneghini2015,
  title={Integrated modeling applications for tokamak experiments with OMFIT},
  author={Meneghini, O and Smith, SP and Lao, LL and Izacard, O and Ren, Q and Park, JM and Candy, J and Wang, Z and Luna, CJ and Izzo, VA and others},
  journal={Nuclear Fusion},
  volume={55},
  number={8},
  pages={083008},
  year={2015},
  publisher={IOP Publishing}
}

@article{Meneghini2020,
  title={Neural-network accelerated coupled core-pedestal simulations with self-consistent transport of impurities and compatible with ITER IMAS},
  author={Meneghini, O and Snoep, Garud and Lyons, Brendan C and McClenaghan, Joseph and Imai, Chieko Sarah and Grierson, B and Smith, Sterling P and Staebler, Gary M and Snyder, Philip B and Candy, Jeff and others},
  journal={Nuclear Fusion},
  volume={61},
  number={2},
  pages={026006},
  year={2020},
  publisher={IOP Publishing}
}

@article{Lyons2023,
  title={Flexible, integrated modeling of tokamak stability, transport, equilibrium, and pedestal physics},
  author={Lyons, BC and McClenaghan, J and Slendebroek, T and Meneghini, O and Neiser, TF and Smith, SP and Weisberg, DB and Belli, EA and Candy, J and Hanson, JM and others},
  journal={arXiv preprint arXiv:2305.09683},
  year={2023}
}

@article{Slendebroek2023,
  title={Elevating zero dimensional global scaling predictions to self-consistent theory-based simulations},
  author={Slendebroek, T and McClenaghan, J and Meneghini, OM and Lyons, BC and Smith, SP and Neiser, TF and Shi, N and Candy, J},
  journal={Physics of Plasmas},
  volume={30},
  number={7},
  year={2023},
  publisher={AIP Publishing}
}

@misc{GAFPP,
    author = {General Atomics},
    title = {Compact fusion power plant},
    year = {2022},
    note = {https://www.ga.com/fusion-pilot-plant}
}

@article{Imbeaux2015,
  title={Design and first applications of the ITER integrated modelling \& analysis suite},
  author={Imbeaux, Fr{\'e}d{\'e}ric and Pinches, SD and Lister, JB and Buravand, Y and Casper, T and Duval, Basil and Guillerminet, B and Hosokawa, Masanari and Houlberg, Wayne and Huynh, Philippe and others},
  journal={Nuclear Fusion},
  volume={55},
  number={12},
  pages={123006},
  year={2015},
  publisher={IOP Publishing}
}

@article{Hawryluk2021,
  title={Bringing Fusion to the US Grid},
  author={Hawryluk, RJ and Garcia-Diaz, BL and Kulcinski, GL and McCarthy, KA and Peterson, PF and Quintenz, JP and Reder, WK and Roop, DW and Snyder, P and Uhle, JL and others},
  journal={The National Academies of Science Engineering and Medicine NASEM Public Briefing February},
  volume={17},
  pages={2021},
  year={2021}
}

@article{Hsu2023,
  title={US Fusion Energy Development via Public-Private Partnerships},
  author={Hsu, Scott C},
  journal={Journal of Fusion Energy},
  volume={42},
  number={1},
  pages={12},
  year={2023},
  publisher={Springer}
}

@article{candy2009,
  title={Tokamak profile prediction using direct gyrokinetic and neoclassical simulation},
  author={Candy, Jeff and Holland, Chris and Waltz, RE and Fahey, Mark R and Belli, E},
  journal={Physics of Plasmas},
  volume={16},
  number={6},
  pages={060704},
  year={2009},
  publisher={American Institute of Physics}
}

@article{belli2008,
  title={Kinetic calculation of neoclassical transport including self-consistent electron and impurity dynamics},
  author={Belli, EA and Candy, J},
  journal={Plasma Physics and Controlled Fusion},
  volume={50},
  number={9},
  pages={095010},
  year={2008},
  publisher={IOP Publishing}
}

@article{neiser2020big,
  title={Big Data Validation of the TGLF Transport Model},
  author={Neiser, Tom and Meneghini, Orso and Smith, Sterling and Fasciana, Michele and Staebler, Gary and Candy, Jeff},
  journal={Bulletin of the American Physical Society},
  volume={65},
  year={2020},
  publisher={APS}
}

@article{Neiser22,
	author = {Tom Neiser and Orso Meneghini and Sterling Smith and Joseph McClenaghan and David Orozco and Joseph Hall and Gary Staebler and Emily Belli and Jeff Candy},
	journal={Bulletin of the American Physical Society},
	title = {Database generation for validation of {TGLF} and retraining of neural network accelerated {TGLF-NN}},
	year = {2022}
}

@article{Staebler07,
	author = {Staebler,G. M. and Kinsey,J. E. and Waltz,R. E.},
	doi = {10.1063/1.2436852},
	journal = {Physics of Plasmas},
	number = {5},
	pages = {055909},
	title = {A theory-based transport model with comprehensive physics},
	volume = {14},
	year = {2007}
}

@article{Staebler16,
	author = {Staebler,G. M. and Candy,J. and Howard,N. T. and Holland,C.},
	doi = {10.1063/1.4954905},
	journal = {Physics of Plasmas},
	number = {6},
	pages = {062518},
	title = {The role of zonal flows in the saturation of multi-scale gyrokinetic turbulence},
	volume = {23},
	year = {2016}
}

@article{Staebler21,
	author = {G.M. Staebler and E. A. Belli and J. Candy and J.E. Kinsey and H. Dudding and B. Patel},
	doi = {10.1088/1741-4326/ac243a},
	journal = {Nuclear Fusion},
	month = {9},
	number = {11},
	pages = {116007},
	publisher = {{IOP} Publishing},
	title = {Verification of a quasi-linear model for gyrokinetic turbulent transport},
	volume = {61},
	year = 2021,
}

@article{meneghini2017,
  title={Self-consistent core-pedestal transport simulations with neural network accelerated models},
  author={Meneghini, Orso and Smith, Sterling P and Snyder, Philip B and Staebler, Gary M and Candy, Jeffrey and Belli, E and Lao, L and Kostuk, Mark and Luce, T and Luda, Teobaldo and others},
  journal={Nuclear Fusion},
  volume={57},
  number={8},
  pages={086034},
  year={2017},
  publisher={IOP Publishing}
}

@article{arbon2020,
  title={Rapidly-convergent flux-surface shape parameterization},
  author={Arbon, Ryan and Candy, Jeff and Belli, Emily A},
  journal={Plasma Physics and Controlled Fusion},
  volume={63},
  number={1},
  pages={012001},
  year={2020},
  publisher={IOP Publishing}
}

@article{lutjens1996,
  title={The CHEASE code for toroidal MHD equilibria},
  author={L{\"u}tjens, Hinrich and Bondeson, Anders and Sauter, Olivier},
  journal={Computer physics communications},
  volume={97},
  number={3},
  pages={219--260},
  year={1996},
  publisher={Elsevier}
}

@article{franza2015,
  title={On the implementation of new technology modules for fusion reactor systems codes},
  author={Franza, F and Boccaccini, LV and Fisher, U and Gade, PV and Heller, R},
  journal={Fusion Engineering and Design},
  volume={98},
  pages={1767--1770},
  year={2015},
  publisher={Elsevier}
}

@article{coleman2019,
  title={BLUEPRINT: a novel approach to fusion reactor design},
  author={Coleman, M and McIntosh, S},
  journal={Fusion Engineering and Design},
  volume={139},
  pages={26--38},
  year={2019},
  publisher={Elsevier}
}

@article{mao2019,
  title={CFETR integration design platform: Overview and recent progress},
  author={Mao, Shifeng and Ye, Minyou and Li, Yang and Zhang, Jianwu and Zhan, Xianda and Wang, Zhongwei and Xu, Kun and Liu, Xufeng and Li, Jiangang},
  journal={Fusion Engineering and Design},
  volume={146},
  pages={1153--1156},
  year={2019},
  publisher={Elsevier}
}

@article{reux2015,
  title={DEMO reactor design using the new modular system code SYCOMORE},
  author={Reux, C{\'e}dric and Di Gallo, Luc and Imbeaux, Fr{\'e}d{\'e}ric and Artaud, J-F and Bernardi, P and Bucalossi, J{\'e}r{\^o}me and Ciraolo, Guido and Duchateau, J-L and Fausser, C and Galassi, Davide and others},
  journal={Nuclear Fusion},
  volume={55},
  number={7},
  pages={073011},
  year={2015},
  publisher={IOP Publishing}
}

@article{franza2022,
  title={MIRA: a multi-physics approach to designing a fusion power plant},
  author={Franza, F and Boccaccini, LV and Fable, E and Landman, I and Maione, IA and Petschanyi, S and Stieglitz, R and Zohm, H},
  journal={Nuclear Fusion},
  volume={62},
  number={7},
  pages={076042},
  year={2022},
  publisher={IOP Publishing}
}

@article{dragojlovic2010,
  title={An advanced computational algorithm for systems analysis of tokamak power plants},
  author={Dragojlovic, Zoran and Raffray, A Rene and Najmabadi, Farrokh and Kessel, Charles and Waganer, Lester and El-Guebaly, Laila and Bromberg, Leslie},
  journal={Fusion Engineering and Design},
  volume={85},
  number={2},
  pages={243--265},
  year={2010},
  publisher={Elsevier}
}

@article{hong2008,
  title={Development of a tokamak reactor system code and its application for concept development of a demo reactor},
  author={Hong, Bong Guen and Lee, Dong Won and Kim, Suk-Kwon and Kim, Yonghee},
  journal={Fusion Engineering and Design},
  volume={83},
  number={10-12},
  pages={1615--1618},
  year={2008},
  publisher={Elsevier}
}

@article{nakamura2012,
  title={Efforts towards improvement of systems codes for the Broader Approach DEMO design},
  author={Nakamura, Makoto and Kemp, Richard and Utoh, Hiroyasu and Ward, David J and Tobita, Kenji and Hiwatari, Ryoji and Federici, Gianfranco},
  journal={Fusion Engineering and Design},
  volume={87},
  number={5-6},
  pages={864--867},
  year={2012},
  publisher={Elsevier}
}

@article{badalassi2023,
  title={FERMI: Fusion Energy Reactor Models Integrator},
  author={Badalassi, V and Sircar, A and Solberg, JM and Bae, JW and Borowiec, K and Huang, P and Smolentsev, S and Peterson, E},
  journal={Fusion Science and Technology},
  volume={79},
  number={3},
  pages={345--379},
  year={2023},
  publisher={Taylor \& Francis}
}

@article{leuer2023,
    author = {James A. Leuer and D. Weisberg and R. MacDonald and I. Favela and P. Beharrell and D. Appelt and R. Buttery and R. Callis and C. Crowe and N. Eidietis and B. Grierson and L. Holland and K. Holtrop and A. Kellman and J. Luxon and C. Murphy and Z. Piec and G. Sip and M. Van Zeeland and A. Zalzali},
    title = {GATM: A 3-D FE Fusion Magnet Model \& Application to DIII-D \& Next Generation Devices},
    journal = {IEEE Transactions on Plasma Science},
    year = {2023},
    volume = {},  % you can fill in the volume if known
    number = {}, % you can fill in the issue number if known
    pages = {},  % you can fill in the page numbers if known
    publisher = {IEEE},
    note = {Member, IEEE},
}

@article{julia,
    title={Julia: A Fresh Approach to Numerical Computing},
    author={Bezanson, Jeff and Edelman, Alan and Karpinski, Stefan and Shah, Viral B.},
    journal={SIAM Review},
    volume={59},
    number={1},
    pages={65--98},
    year={2017},
    publisher={SIAM},
    doi={10.1137/141000671},
    url={https://julialang.org/publications/julia-fresh-approach-BEKS.pdf}
}

@article{sheffield2016, 
    title={Generic Magnetic Fusion Reactor Revisited},
    author={John Sheffield and Stanley L. Milora},
    journal={Fusion Science and Technology},
    volume={70},
    number={1},
    pages={14--35},
    year={2016},
    doi={10.13182/FST15-157}
}

@article{sheffield1986,
	author = {J. Sheffield and R. A. Dory and S. M. Cohn and J. G. Delene and L. Parsly and D. E. T. F. Ashby and W. T. Reiersen},
	doi = {10.13182/FST9-2-199},
	eprint = {https://doi.org/10.13182/FST9-2-199},
	journal = {Fusion Technology},
	number = {2},
	pages = {199-249},
	publisher = {Taylor & Francis},
	title = {Cost Assessment of a Generic Magnetic Fusion Reactor},
	url = {https://doi.org/10.13182/FST9-2-199},
	volume = {9},
	year = {1986}}

@techreport{waganer2013,
  title={ARIES Cost Account Documentation},
  author={L.M. Waganer},
  year={2013},
  institution={University of California, San Diego}
}

@techreport{lengyel1981,
  title={Analysis of radiating plasma boundary layers},
  author={Lengyel, LL},
  year={1981},
  institution={Max-Planck-Institut f{\"u}r Plasmaphysik}
}

@article{metaheuristics2022, 
  doi = {10.21105/joss.04723}, 
  url = {https://doi.org/10.21105/joss.04723}, 
  year = {2022}, 
  publisher = {The Open Journal}, 
  volume = {7}, 
  number = {78}, 
  pages = {4723}, 
  author = {Jesús-Adolfo Mejía-de-Dios and Efrén Mezura-Montes}, 
  title = {Metaheuristics: A Julia Package for Single- and Multi-Objective Optimization}, 
 journal = {Journal of Open Source Software} }

@article{humphreys2008,
  title={DIII-D integrated plasma control solutions for ITER and next-generation tokamaks},
  author={Humphreys, DA and Ferron, JR and Hyatt, AW and La Haye, RJ and Leuer, JA and Penaflor, BG and Walker, ML and Welander, AS and In, Yongkyoon},
  journal={Fusion engineering and design},
  volume={83},
  number={2-3},
  pages={193--197},
  year={2008},
  publisher={Elsevier}
}

@inproceedings{grierson2023,
  title={Design and Technology Maturation of General Atomics Steady-State Advanced Tokamak Fusion Pilot Plant},
  author={Grierson, B.A and Solomon, W.M. and Weisberg, D. and Meneghini, O. and Humphreys, D. and Akiyama, T. and Leuer, J. and Eidietis, N. and Abrams, T. and Tillack, M. and Beharrell, P. and Zeller, K. and Mijatovic, P. and MacDonald, R. and Favela, I. and McLaughlin, K. and Xing, A. and Shi, N. and Guterl, J. and Yu, J. and Harvey, J. and Raffray, R. and Appelt, D.},
  booktitle={Proceedings of the IAEA FEC 2023 Conference},
  year={2023}
}

@inproceedings{weisberg2023b,
  title={Integrated Design and Optimization of the Advanced Tokamak Path Toward the Steady-State Fusion Pilot Plant},
  author={Weisberg, D.B. and Grierson, B.A. and Leuer, J. and Shi, N. and Meneghini, O. and Zeller, K. and Guterl, J. and Akiyama, T. and Beharrell, P. and Cote, T. and Ding, S. and Favela, I. and MacDonald, R. and McLaughlin, K. and Mijatovic, P. and Raffray, R. and Solomon, W. and Tillack, M.},
  booktitle={Proceedings of the IAEA FEC 2023 Conference},
  year={2023}
}

@article{muldrew2024,
  title={Conceptual design workflow for the STEP Prototype Powerplant},
  author={Muldrew, Stuart I and Harrington, Chris and Keep, Jonathan and Waldon, Chris and Ashe, Christopher and Chapman, Rhian and Griesel, Charles and Pearce, Alexander J and Casson, Francis and Marsden, Stephen P and others},
  journal={Fusion Engineering and Design},
  volume={201},
  pages={114238},
  year={2024},
  publisher={Elsevier}
}

@article{portone2005,
  title={The stability margin of elongated plasmas},
  author={Portone, Alfredo},
  journal={Nuclear fusion},
  volume={45},
  number={8},
  pages={926},
  year={2005},
  publisher={IOP Publishing}
}

@article{humphreys2009,
  title={Experimental vertical stability studies for ITER performance and design guidance},
  author={Humphreys, DA and Casper, TA and Eidietis, N and Ferrara, M and Gates, DA and Hutchinson, IH and Jackson, GL and Kolemen, Egemen and Leuer, JA and Lister, J and others},
  journal={Nuclear Fusion},
  volume={49},
  number={11},
  pages={115003},
  year={2009},
  publisher={IOP Publishing}
}

@misc{julia_benchmarks,
  title = {Julia Language Benchmarks},
  howpublished = {\url{https://julialang.org/benchmarks/}}
}

@misc{ptp,
  title = {Project Torrey Pines},
  howpublished = {\url{https://github.com/ProjectTorreyPines}}
}

@misc{fuse_help,
  title = {FUSE},
  howpublished = {\url{https://fuse.help}}
}

@misc{apache2,
  author       = {Apache Software Foundation},
  title        = {Apache License, Version 2.0},
  year         = {2004},
  url          = {https://www.apache.org/licenses/LICENSE-2.0}
}

@misc{transp,
  author       = {Princeton Plasma Physics Laboratory},
  title        = {TRANSP Code},
  howpublished = {Computer Software},
  year         = {2024},
  url          = {https://transp.pppl.gov/},
  doi          = {10.11578/dc.20180627.4}
}

@techreport{onetwo,
  title       = {ONETWO: A Computer Code for Modeling Plasma Transport in Tokamaks},
  author      = {Pfeiffer, W. W. and Davidson, R. H. and Miller, R. L. and Waltz, R. E.},
  institution = {General Atomics},
  year        = {1980},
  number      = {GA-A16178},
  doi         = {10.2172/6861782},
  url         = {https://www.osti.gov/biblio/6861782},
  month       = {12}
}

@article{olofsson2022,
  title   = {Vertical instability growth rate studies with rigid and deformable plasma models and proximity controller development in the TCV tokamak},
  author  = {Olofsson, K. E. and other authors},
  journal = {Plasma Physics and Controlled Fusion},
  volume  = {64},
  year    = {2022},
  pages   = {072001},
  doi     = {10.1088/1361-6587/ac7f14},
}

@article{weiland2018,
  title={RABBIT: Real-time simulation of the NBI fast-ion distribution},
  author={Weiland, M and Bilato, R and Dux, R and Geiger, B and Lebschy, A and Felici, F and Fischer, R and Rittich, D and Van Zeeland, M and ASDEX Upgrade Team and others},
  journal={Nuclear Fusion},
  volume={58},
  number={8},
  pages={082032},
  year={2018},
  publisher={IOP Publishing}
}

@article{pankin2004,
  title={The tokamak Monte Carlo fast ion module NUBEAM in the National Transport Code Collaboration library},
  author={Pankin, Alexei and McCune, Douglas and Andre, Robert and Bateman, Glenn and Kritz, Arnold},
  journal={Computer Physics Communications},
  volume={159},
  number={3},
  pages={157--184},
  year={2004},
  publisher={Elsevier}
}

@article{mcclenaghan2023,
  title={Self-consistent investigation of density fueling needs on ITER and CFETR utilizing the new Pellet Ablation Module},
  author={McClenaghan, J and Lao, LL and Parks, PB and Wu, W and Zhang, J and Chan, VS},
  journal={Nuclear Fusion},
  volume={63},
  number={3},
  pages={036015},
  year={2023},
  publisher={IOP Publishing}
}

@inproceedings{neiser2024,
  author       = {T. F. Neiser and D. Sun and B. Agnew and T. Slendebroek and O. Meneghini and B. C. Lyons and A. Ghiozzi and J. McClenaghan and G. Staebler and J. Candy},
  title        = {TJLF: The quasi-linear model of gyrokinetic transport TGLF translated to Julia},
  booktitle    = {Proceedings of the 2024 APS-DPP Annual Meeting},
  year         = {2024},
  address      = {San Diego, CA},
  publisher    = {American Physical Society},
  url          = {https://julialang.org},
}

@inproceedings{walker2015,
  title={Development environments for Tokamak plasma control},
  author={Walker, ML and Humphreys, DA and Sammuli, B and Welander, AS and Winter, A and Snipes, J and de Vries, P and Ambrosino, Giuseppe and De Tommasi, Gianmaria and Mattei, Massimiliano and others},
  booktitle={2015 IEEE 26th Symposium on Fusion Engineering (SOFE)},
  pages={1--8},
  year={2015},
  organization={IEEE}
}

@article{romano2015,
  title={OpenMC: A state-of-the-art Monte Carlo code for research and development},
  author={Romano, Paul K and Horelik, Nicholas E and Herman, Bryan R and Nelson, Adam G and Forget, Benoit and Smith, Kord},
  journal={Annals of Nuclear Energy},
  volume={82},
  pages={90--97},
  year={2015},
  publisher={Elsevier}
}

@article{cerfon2010,
    author = {Cerfon, Antoine J. and Freidberg, Jeffrey P.},
    title = "{“One size fits all” analytic solutions to the Grad–Shafranov equation}",
    journal = {Phys. Plasmas},
    volume = {17},
    number = {3},
    pages = {032502},
    year = {2010},
    month = {03},
    issn = {1070-664X},
    doi = {10.1063/1.3328818},
    url = {https://doi.org/10.1063/1.3328818},
    eprint = {https://pubs.aip.org/aip/pop/article-pdf/doi/10.1063/1.3328818/14817839/032502\_1\_online.pdf},
}

@article{titus2013,
  title={DEsign and analysis of the iter vertical stability (vs) coils},
  author={Titus, Peter H and Kalish, Michael and Hause, Christopher M and Heitzenroeder, Philip and Hsiao, Jushin and Pillsbury, Robert and Daly, Edward},
  journal={Fusion Science and Technology},
  volume={64},
  number={2},
  pages={136--145},
  year={2013},
  publisher={Taylor \& Francis}
}

@article{sugama1998,
  title={Nonlinear electromagnetic gyrokinetic equation for plasmas with large mean flows},
  author={Sugama, Hideo and Horton, W},
  journal={Physics of Plasmas},
  volume={5},
  number={7},
  pages={2560--2573},
  year={1998},
  publisher={American Institute of Physics}
}

@misc{GAMFE,
  title        = {General Atomics Magnetic Fusion},
  author       = {{General Atomics}},
  url          = {https://www.ga.com/magnetic-fusion/}
}

\end{document}